\documentclass[journal]{IEEEtran}

\usepackage[numbers]{natbib}
\usepackage{url}
\usepackage{amsmath}
\usepackage{amssymb}
\usepackage{mathtools}
\usepackage{enumitem}
\usepackage{graphicx}
\usepackage{subcaption}

\begin{document}

\title{Unified Diagnostics for Quantifying AC Operating-Point Robustness Under Injection and Topological Uncertainties with Regime Changes}

\author{Lauren\c{t}iu~Lucian~Anton*$^{1}$, %
    \thanks{* Corresponding author.} %
    \thanks{$^{1}$ ORCID: 0000-0001-7129-2256}%
\and
Marija~Ili\'c$^{\,2}$%
\thanks{$^{2}$ ORCID: 0000-0002-6835-197X}%
\\ \emph{Massachusetts Institute of Technology, 
77 Massachusetts Ave, Cambridge, MA 02139, United States}}

\markboth{Draft for journal submission, Feb 2026}%
{Draft for journal submission, Feb 2026}

\maketitle

\begin{abstract}
    In the presence of uncertainties in load, generation, and network topology, power system planning must reflect operational conditions, while operations require situational awareness over credible uncertainty sets. Diverse methods exist to screen, analyze, embed, and propagate operational uncertainty in power flow and optimal power flow settings. While complementary, these approaches offer only partial insight into how physical constraints, controls, and economic interactions shape the robustness of steady-state operating-points under uncertainty. By formulating operating-point robustness as a post-solution physical response problem around a solved AC optimal power flow (AC-OPF) equilibrium, this paper provides a unified framework for assessing robustness under injection and topological uncertainty without re-optimization. We construct a primal physical response mapping that accounts for connectivity changes, active power redistribution, generator saturation including $PV\!\rightarrow\!PQ$ transitions, and AC network propagation, and introduce quasi-duals that offer a geometric interpretation of shadow prices for off-optimal equilibria. Using these response mappings, we develop a suite of deterministic screening procedures that generalize $N\!-\!k$ contingency analysis to include cost vulnerability $C\!-\!k$, and local analogs $N\!+\!\delta(k)$ and $C\!+\!\delta(k)$, defined through sensitivity-normalized margins and risk tolerances. We extend the framework to include probabilistic screening for distribution-based and moment-based uncertainties, and introduce sequentially-pruned mixture modeling and $\alpha$-stressed regime constructions to manage combinatorial branching from probabilistic regime transitions. We present a case study on the Puerto Rican bulk power system that demonstrates how violation and vulnerability diagnostics can be integrated with geospatial data to produce visualizations that enhance situational awareness for operational and planning applications.
\end{abstract}


\section{Introduction}
\label{sec:intro}
Modern power systems operate under growing uncertainty, jointly driven by imperatives to decarbonize generation, electrify end-use, and maximize network utilization. As variable renewable capacity expands, dispatchable thermal capacity continues to retire. These fleet changes, shaped by both economics and policy incentives, vary by technology and region and can reduce flexibility in balancing supply and demand \cite{2024_iea_weo, 2024_irena_renewable_costs, 2023_nerc_ibr}. At the same time, electrification of heating and transport, combined with rapid data-center and industrial growth, is contributing to sustained demand increases across sectors \cite{2024_review_net_zero, 2025_nyiso_powertrends}. Transmission and distribution networks, already under economic and regulatory pressure, are now operated with reduced margins while increasingly exposed to climate-related risks \cite{2021_iea_climate_resilience}.

Under these stressors, even small differences between modeled and realized operating conditions, arising from forecast errors, modeling inaccuracies, deviations in generator or DER output, or topology changes, can push a system toward thermal, voltage, or stability limits. This motivates not only renewed interest in constraint margins and uncertainty propagation, but also a systematic view of operating-point robustness to realistic uncertainties in injections and network parameters. A diverse set of fast screening tools exists, each addressing specific uncertainties but providing only partial visibility into how likely an operating point remains within limits over different uncertainty sets. Traditional security criteria, which perform systematic screening of robustness, are formulated in terms of discrete disturbance sets such as $N\!-\!1$ and selected $N\!-\!k$ contingencies, and most probabilistic or risk-based extensions remain tied to these contingency lists. Given the heterogeneous uncertainties modern power systems must contend with, operators and planners benefit from lightweight, interpretable diagnostics that generalize these notions across varied uncertainty sets, whether discrete, local, or probabilistic.

\subsection{Background}
\label{sec:intro:back}

We substantiate this narrative in detail. We first highlight recent operator reports that document operational trends affecting constraint margins, motivating renewed attention to operating-point robustness. We then provide a high-level overview of analytical tools developed to capture the effects of injection and topological uncertainty on steady-state equilibria, both to summarize the state of the art and to motivate a unified framework that builds on these mature foundations to deliver new system-wide diagnostics of operating-point robustness over meaningful uncertainty sets.

\vspace{0.25cm}
\subsubsection{Operational Trends on Constraint Margins}
\label{sec:intro:back:trends}
Across the United States, reports from system operators speak to the combined pressures of increasing uncertainties in generation capabilities, load profiles, and system states, while operating with reduced margins. 

NY-ISO warns that as more fossil generators retire, new carbon-free resources are not being added fast enough to keep pace with expected demand growth, raising concerns about whether the aging fleet will be able to provide essential grid reliability services under shrinking margins \cite{2025_nyiso_powertrends}. ISO-NE reports ``dramatic changes'' in its energy mix, shifting from coal and oil to natural gas, with an interconnection queue dominated by wind and batteries. With steady growth expected from increased heating and transport electrification, a shift from summer to winter peaking patterns brings concerns over energy adequacy and deliverability on both the grid and gas pipeline systems \cite{2024_isone_grid_outlook, 2024_isone_reo}.

Both PJM and MISO caution that dispatchable resources are increasingly being replaced by weather-dependent generation with materially different capabilities, raising concerns over operational challenges associated with limited-duration storage, extreme-weather exposure, and large single-site load additions \cite{2023_pjm_energy_transition, 2024_miso_reliability_imperative}. CAISO reports growing mismatches between supply and demand resulting in wind and solar curtailment, increased ramping requirements, mentions heightened climate-related threats to grid infrastructure, and that battery storage now plays an important role in power balancing \cite{2025_caiso_evolving_grid}.

ERCOT highlights significant projected load growth, increased thermal generation retirement, rapid growth in both transmission-connected and distributed wind, solar, and energy storage development \cite{2024_ercot_constraints_needs}. SPP similarly comments on shrinking reserve margins that are expected to fall below its 15\% Planning Reserve Margin by 2027, resulting from thermal retirements and expiring contracts, while expressing an increased reliance on demand response  \cite{2025_spp_summer_resource_adequacy}. 

Puerto Rico represents perhaps the most acute example of these operational challenges. Operated by LUMA Energy, the grid remains fragile, contending with compounding climate-related damages, while simultaneously working towards rapid grid decarbonization and modernization. Recurrent blackouts, limited flexible reserves, narrow operating margins, and funding constraints continue to impact progress towards reliable, resilient operation \cite{2025_luma_q4_metrics_outages, 2025_luma_resource_adequacy_report}.

The reports above collectively point to increasing reliability concerns and growing vulnerability to uncertain realizations in generation, demand, or topology under tighter operating margins. These trends motivate renewed attention to how close an operating point lies to its physical, regulatory, and stability limits, and how quickly such margins are consumed under uncertainty. We next survey the body of analytical methods that characterize how injection and topological perturbations propagate through steady-state network equations.

\vspace{0.25cm}
\subsubsection{Approaches to Incorporating Uncertainty in Steady-State Power Flow and Dispatch}
\label{sec:intro:litrev}
Operational uncertainty has long been addressed through methods that either embed uncertainty directly into dispatch formulations or provide post-solution assessments of how perturbations propagate through steady-state network equations. These two classes form the foundation of existing notions of operating-point robustness.

A substantial body of work incorporates uncertainty directly into optimal power flow (OPF) formulations. Stochastic, chance-constrained, robust, and distributionally robust OPF treat injections or model parameters as random variables or uncertainty sets, enforcing operational limits with a specified confidence level or across worst-case realizations. Chance-constrained OPF (CC-OPF) requires that constraints hold with probability at least $1-\epsilon$ \cite{2014_bienstock_ccopf}, while stochastic OPF (S-OPF) often constrains or optimizes a conditional value-at-risk (CVaR). For topological uncertainty, robust OPF (R-OPF) enforces feasibility across specified contingency sets, and distributionally robust OPF (DR-OPF) optimizes performance under the worst-case distribution within a defined uncertainty set \cite{2023_dropf, 2018_dropf_data}. In practice, market processes such as security-constrained unit commitment and economic dispatch (SC-UC and SC-ED) model contingencies and reserve requirements to handle uncertainty prior to real-time operation. These approaches provide ex ante guarantees by embedding uncertainty, rather than assessing the robustness of a specific realized AC operating point in post.

Despite the variety of tools for embedding uncertainty in dispatch, realized system states often occur that were not considered in predictive formulations. A complementary suite of methods focus on post-solution screening of operating points, to specific contingencies. Screening for likely post-contingency constraint violations is often performed using linearized distribution factors that consider incremental changes in power transfers (PTDF), line outages (LODF), line additions (LADF), and outage transfers (OTDF) \cite{2015_powerworld_linear_analysis}. These methods provide local sensitivity information, treat uncertainty sources one at a time, and often rely on linear approximations that are accurate within a limited neighborhood of the base case.

Probabilistic load flow (PLF) methods propagate uncertainty in injections or parameters to state variables such as voltages and branch flows. Monte Carlo sampling estimates violation frequencies by repeatedly solving AC power flows under sampled conditions \cite{1985_plf_network_outages}. Analytical PLF approaches propagate input moments or cumulants through linearized mappings to approximate state probability distributions \cite{1974_plf_first, 1975_stochastic_load_flows, 2009_pfl_with_wind}. Point estimation methods (PEMs) approximate these distributions using a small number of deterministic points derived from input moments \cite{2017_pfl_point_estimation}. PLF has also been formulated to consider constrained load flow and corrective control actions to minimize violations \cite{1994_nikos_pclf}.

Robustness to steady-state voltage stability limits is commonly assessed through homotopy-based continuation power flow (CPF), which traces loadability curves and identifies the bifurcation point that defines maximum loadability \cite{1992_cpf}, beyond which voltage collapse can occur. Steady-state stability margins are thus defined as a distance to this point. In contrast, PLF extensions have used linearized sensitivities to estimate the probability of collapse and identify corrective control actions \cite{1998_nikos_plf_voltage}. These methods each provide important insights into voltage phenomena but focus on a specific degrees of freedom rather than broad robustness assessments.

A related class of methods considers risk-based security assessment, where the probability of a contingency and the severity of its consequences are combined to compute expected risk indices \cite{1996_billinton_book}. These methods can be interpreted as probabilistic generalizations of $N\!-\!1$ screening, since the underlying uncertainty sets are enumerated contingency scenarios rather than continuous injection or topology uncertainty models. As such, risk-based approaches extend deterministic formulations, rather than providing a uniform PFL-style treatment of uncertainty around defined operating points.

Taken together, these approaches offer valuable yet fragmented notions of operating-point robustness, each defined with respect to particular physical mappings and uncertainty sets that differ substantially in structure, dimensionality, and operational meaning. None provide a unified way to assess how a solved AC-OPF equilibrium responds to heterogeneous injection and topology uncertainty. As a result, existing notions of robustness remain dispersed across tools developed for different purposes, making it difficult to interpret what it means for an operating point to be ``robust,'' and with respect to which uncertainties. This motivates the development of a unified post-solution framework that evaluates robustness with respect to clearly defined and meaningful uncertainty sets around an operating point.

\vspace{0.25cm}
\subsubsection{Toward a Unified Framework for Operating-Point Robustness}
\label{sec:intro:unify}

A unified view of operating-point robustness can be defined relative to a known network state and a specified uncertainty set, and should consider both physical feasibility and optimality. An AC optimal power flow (AC-OPF) solution provides such a reference, as it specifies a feasible AC operating point that satisfies nonlinear power balance equations and operational limits while optimizing a chosen objective, often cost. The associated primal and dual information jointly characterize physical margins and economic tightness, making AC-OPF a natural generalized reference formulation for propagating injection and topological perturbations in a unified robustness framework.

While an AC-OPF solution fully specifies base operating points, it does not by itself define how the system responds to physical perturbations. Changes in steady-state equilibria should reflect layered active and reactive power control mechanisms, on the timescale of interest. AC power flow (AC-PF) and many PLF formulations assume that active power imbalances are absorbed by a single slack bus \cite{2008_plf_review}, with distributed slack treatments considered in specific settings \cite{2013_plf_distributed}. These are typically interpreted as feedback control to frequency deviations, as in automatic generation control (AGC), in contrast to feed-forward preventive or proportional reserve allocations that balance predicted imbalances. Reactive power control introduces further complexities through control regime transitions for voltage-controlling generators when reactive limits are reached. These saturation effects are critical for realistic steady-state voltage behavior but poorly represented in PLF studies \cite{2004_plf_pvpq}. A unified robustness framework must therefore encode these active and reactive control layers, rather than relying solely on a single slack-bus or static bus types.

Perturbations applied to an AC-OPF equilibrium with specified control logic generally drive the system toward a new steady-state that is no longer optimal with respect to the original objective and may violate operational limits. In this \emph{off-optimal} regime, the Karush–Kuhn–Tucker (KKT) conditions no longer hold and an exact dual formulation no longer exists. Nevertheless, local projections of constraint sensitivities can be interpreted as \emph{quasi-dual} signals that approximate how tight constraints are economically in off-optimal conditions, enabling robustness assessment from both primal and quasi-dual perspectives.

Once a physical response mapping is specified, operating-point robustness can be assessed with respect to arbitrary perturbations in injections and topology. The various notions of robustness reviewed in Section~\ref{sec:intro:litrev} consider different uncertainty sets and a base operating point. A unified framework should therefore admit these diverse uncertainty sets within a common formulation, so that conventional criteria like $N\!-\!1$ security become special cases.

In many practical settings, detailed probability distributions for uncertain injections or parameters are unavailable, and only limited statistical information can be estimated reliably. In such cases, distribution-free inequalities such as Cantelli bounds provide conservative guarantees on constraint violation probabilities based solely on means and variances. These bounds have been used to derive tractable reformulations in CC-OPF problems \cite{2020_ccopf_cantelli}, but, to the best of our knowledge, have not been adopted as post-solution diagnostics for AC-OPF operating points, using PLF techniques. A unified robustness framework should accommodate both distributional models and moment-based bounds, enabling conservative robustness assessments when uncertainty information is incomplete.

A unified formulation should thus generalize traditional security notions by quantifying exposure to constraint violations, erosion of margins, and potential constraint release under deterministic or probabilistic perturbations. In practice, these diagnostic layers can be aggregated into system-level visualizations that summarize robustness across entire uncertainty sets, extending traditional single-contingency views into multi-scenario assessments. Such visualizations support situational awareness by revealing spatial patterns in state margins and violations, relevant to both operational and planning contexts.

\subsection{Contributions}
\label{sec:intro:cont}

This paper develops a unified post-solution framework for assessing the robustness of AC-OPF operating points under injection and topology uncertainties. In doing so, we:

\begin{itemize}[itemsep=0pt, topsep=2pt]
    \item Formalize operating-point robustness of a solved AC-OPF equilibrium by unifying deterministic contingency analysis, local perturbation screening, and probabilistic uncertainty modeling within a common framework.

    \item Develop a control- and constraint-aware steady-state response mapping that incorporates regime changes resulting from islanding, generator control layers, and saturation effects when propagating injection and topology perturbations through the network.

    \item Propose a quasi-dual construction that extends the interpretation of shadow prices to off-optimal steady-states reached under small perturbations.

    \item Define deterministic and probabilistic robustness metrics for post-solution screening, including margin-based indicators, violation and release probabilities, and moment-based bounds for when uncertainty information is limited.

    \item Introduce probabilistic methods for handling combinatorial, uncertainty-induced regime changes, including a sequentially pruned mixture model for distributional uncertainty and an $\alpha$-stressed construction for moment-based uncertainty.

    \item Develop system-wide diagnostic layers that unify primal, quasi-dual, and probabilistic information, enabling interpretable visual assessments of robustness over specified uncertainty sets.

    \item Demonstrate the framework on the Puerto Rican bulk power system, illustrating its ability to visualize system-wide vulnerabilities.
\end{itemize}

The rest of the paper is organized as follows. Section~\ref{sec:problem} develops the framework for steady-state AC physical response. Sections~\ref{sec:deterministic} and~\ref{sec:probabilistic} introduce the proposed deterministic and probabilistic robustness screening tools. Section~\ref{sec:casestudy} demonstrates these tools on the Puerto Rican bulk power system, with visualizations. Section~\ref{sec:conclusion} summarizes the findings and outlines directions for future work.

\section{Steady-State Physical Response to Uncertainty Around AC-OPF Equilibria}
\label{sec:problem}

Assessing robustness requires understanding how an AC system in equilibrium responds to small changes in injections or network parameters. An AC-OPF solution provides an operating point that is optimal over the feasible region defined by the constraint set. Primal variables encompass bus voltages, branch flows, and generator dispatch, equality constraints determine the feasible manifold and inequality constraints determine the feasible region(s) on that manifold.

To quantify how constraint margins are affected, we model the steady-state response of the system to variations in active and reactive injections and in network admittances. From an AC-OPF equilibrium, we construct a primal perturbation mapping that considers active power redistribution via participation factors and generator limits, reactive power saturation via $PV\rightarrow PQ$ regime changes, and propagation through AC network equations around the base point. We then introduce a quasi-dual construction that interprets how the cost gradient aligns with the local constraint manifold when the system settles at a physically feasible but off-optimal equilibrium.

\subsection{AC-OPF Primal Formulation and Notation}
\label{sec:problem:primal}

Without loss of generality, we consider an AC-OPF formulation for active power economic dispatch. Let $\mathcal{N}$, $\mathcal{G}$, $\mathcal{D}$, and $\mathcal{B}$ be the sets of buses, generators, loads, and branches, respectively. Let $\mathbf{Y}_{\mathrm{bus}}$ be the admittance matrix, defined by considering all network components including series and shunt compensation. $\mathbf{Y}_{\mathrm{bus}}$ has complex elements $Y_{ij} = |Y_{ij}|\angle\delta_{ij}=G_{ij}\!+\!jB_{ij}$ between buses $i$ and $j$, that can be represented either using admittance magnitude $|Y_{ij}|$ and angle $\delta_{ij}$, or conductance $G_{ij}$ and susceptance $B_{ij}$. Define apparent power as $S\!=\!P\!+\!jQ$, with active power $P$, and reactive power $Q$, and bus voltage $V=|V|\angle\theta$, with voltage magnitude $|V|$ and angle $\theta$. At each bus $i$, define $P_i=\sum_{g\in\mathcal{G}_i}P_g - \sum_{d\in\mathcal{D}_i}P_d$ and $Q_i=\sum_{g\in\mathcal{G}_i}Q_g - \sum_{d\in\mathcal{D}_i}Q_d$ as the net active and reactive power injection given generators $\mathcal{G}_i \subseteq \mathcal{G}$ and loads $\mathcal{D}_i \subseteq \mathcal{D}$ connected at bus $i$. Let $S_{ij}=V_i(V_i\!-\!V_j)^*Y_{ij}^*$ and $\theta_{ij}=\theta_i\!-\!\theta_j$ be the apparent power flow and angle difference across a branch connecting buses $i$ and $j$. Branches are taken to be either lines or transformers. Lastly, assume at most quadratic functions $f(P_g) = a_g P_g^2 + b_g P_g + c_g$ in the objective function. We then have:
\begin{align}
    & \!\!\!\!\!\!\!\!\!\!\!\!\!\!\! \min_{\{P_g,\, Q_g,\, |V_i|,\, \theta_i\}} \,\,
         \sum_{g\in\mathcal{G}} f(P_g) \label{eq:primal_objective} \\
    \mathrm{s.t.} \,\,  
    & P_i \!=\! |V_i| \sum_{j\in\mathcal{N}} |Y_{ij}| |V_j| \cos(\theta_{ij} \!-\! \delta_{ij}) 
        \,\, \forall \,i \!\in\! \mathcal{N} \label{eq:P_bal} \\
    & Q_i \!=\! |V_i| \sum_{j\in\mathcal{N}} |Y_{ij}| |V_j| \sin(\theta_{ij} \!-\! \delta_{ij}) 
        \,\, \forall \,i \!\in\! \mathcal{N} \label{eq:Q_bal} \\
    & P_g^{\min} \le P_g \le P_g^{\max}, 
        \,\, \forall \,g \!\in\! \mathcal{G} \label{eq:P_lims} \\
    & Q_g^{\min} \le Q_g \le Q_g^{\max}, 
        \,\, \forall \,g \!\in\! \mathcal{G} \label{eq:Q_lims} \\
    & V_i^{\min} \le |V_i| \le V_i^{\max}, 
        \,\, \forall \,i \!\in\! \mathcal{N} \label{eq:V_lims} \\
    & \,\, |S_{ij}| \le S_{ij}^{\max}, 
        \,\, \forall \,(i,j) \!\in\! \mathcal{B} \label{eq:flow_lims} \\
    & \,\,\, \theta_{ij}^{\min} \le \theta_{ij} \le \theta_{ij}^{\max}, 
        \,\, \forall \,(i,j) \!\in\! \mathcal{B} \label{eq:ang_lims} \\
    & \quad\, \theta_{i_0} = \, 0, 
        \,\, \text{for the reference bus.} \label{eq:ang_ref}
\end{align}

We represent \eqref{eq:P_bal} and \eqref{eq:Q_bal} with $|Y_{ij}|$ and $\angle\delta_{ij}$ instead of $G_{ij}$ and $B_{ij}$, for compactness. Fixed voltage setpoints can be modeled with tight box limits in \eqref{eq:V_lims} or replaced with equality constraints $V_i=V_i^{\mathrm{set}}$. Branch limits may constrain $P_{ij}$ rather than $S_{ij}$. We lastly note that many other objective functions are possible for AC-OPF, including those that assign costs for reactive power compensation or emissions \cite{2023_acopf_review}.

\subsection{Modeling Steady-State AC Physical Response}
\label{sec:problem:response}

How a system responds to changes in inputs depends on the timescale of interest. Here we assume a continuous and stable transition between steady-state AC equilibria, without triggering protection or discrete switching events.

We model the physical response of a system in two stages. First we model distributed active power balancing, to generate a combined disturbance vector with a minimized imbalance. This combined response is then used to parameterize a standard AC-PF problem. Together, these stages characterize the physical response of the network using both generator controls and power balance equations \eqref{eq:P_bal}--\eqref{eq:Q_bal}.

\vspace{0.25cm}
\subsubsection{AC Power Flow Bus Conventions}
\label{sec:problem:response:buses}
In traditional AC-PF, two of four bus variables ($P$, $Q$, $|V|$, and $\theta$) are assigned by a given dispatch, thereby labeling the buses by what is fixed. The standard bus definitions are $PV$-buses, where generator active power output is fixed and the bus voltage is locally controlled via generator reactive power output; $PQ$-buses, where the active and reactive power injections are assigned fixed, often at load buses; and a slack bus, which both serves as a voltage reference bus, and also the reserve generator bus that absorbs all mismatches after AC-PF is solved for all other buses \cite{2012_overbye_book}.

\vspace{0.25cm}
\subsubsection{Fixed and Non-Controllable Elements}
\label{sec:problem:response:loads}
We treat all loads as fixed $PQ$ elements. Non-dispatchable generation (e.g., renewable resources without active control) or generation not participating in active power balancing are represented as negative loads. Thus, buses containing only fixed injections are modeled as $PQ$-buses.

\vspace{0.25cm}
\subsubsection{Generator Active Power Response}
\label{sec:problem:response:active}
When defining a PF problem from an OPF state, active power redistribution must be defined. We distinguish four power balancing conventions:
\begin{enumerate}[itemsep=0pt, topsep=2pt]
    \item \emph{Single-slack:} a simplified representation, where one large generator absorbs all power mismatches;
    \item \emph{Primary control:} distributed compensation over participating generators via frequency-droop response;
    \item \emph{Secondary control:} distributed compensation via automatic generator control (AGC) response, acting on slower timescales and restoring nominal frequency;
    \item \emph{Tertiary control:} redispatch via a new OPF solve, representing an economic re-optimization rather than a physical response, and not considered here.
\end{enumerate}

The single-slack formulation is the standard AC-PF representation, with a designated slack generator at the reference bus absorbing active power imbalances. Both primary and secondary frequency control admit a distributed slack interpretation using participation factors. Primary control acts over seconds via governor power–frequency droop to restore power balance, while secondary control acts over minutes via AGC to restore nominal frequency and eliminate area control error (ACE) \cite{1994_kundur,2000_ilic_zaborszky}. In both cases, the resulting steady-state redistribution corresponds to automated feedback in response to frequency deviations, rather than preplanned, feed-forward corrective redispatch conditioned on specific disturbance scenarios. Apart from how participation factors are determined, these control layers are algebraically equivalent at equilibrium and define the active power response model adopted here.

We model all controllable generators as responsive elements with hard capacity limits on active and reactive power, represented in \eqref{eq:P_lims} and \eqref{eq:Q_lims}. Let $\mathcal{G}_r\subseteq\mathcal{G}$ be the set of dispatchable generators participating in automated active power balancing. We consider the generalized case, where more than one generator can participate in active power balance via participation factors, $\alpha_g$, where  $\sum_{g\in\mathcal{G}_r}\alpha_g=1$. Generators whose injections are directly perturbed are excluded from the balancing set $\mathcal{G}_r$.

Assume a net perturbation in active power injections, $dP_{\mathrm{inj}}=\sum_{i\in\mathcal{N}_p}dP_i$ where $\mathcal{N}_p$ is the set of buses with perturbed injections. The response from each participating generator is then
\begin{equation}
    dP_g = -\alpha_gdP_{\mathrm{inj}}. 
    \label{eq:ideal_redistribution}
\end{equation}
This defines a local feasibility range for perturbations, or necessitates at least one generator without hard constraint limits. Further, the response from each participating generator should, in principle, respond to the total imbalance including incremental losses, $dP_{\mathrm{losses}}$. However, since the incremental losses are not known a~priori, these must be implicitly handled. Here, we assume that the incremental losses are picked up by a designated generator $g_0$ after $dP_{\mathrm{inj}}$ is distributed among participating generators. If $g_0$ is at the reference bus, this coincides with the numerical slack bus in AC-PF formulations.

\vspace{0.25cm}
\paragraph{Generator Limit Considerations}
We consider active power operational limits using an iterative residual-imbalance method. Let $dP_{\mathrm{res}}^{(\ell)}$ denote the residual net-injection imbalance at iteration $\ell$, initialized as $dP_{\mathrm{res}}^{(0)} = dP_{\mathrm{inj}}$. For each participating generator $g \in \mathcal{G}_r^{(\ell)}$, the predicted response is
\begin{equation}
    P_g^{\mathrm{pred},(\ell)} = P_g + dP_g^{(\ell)} = P_g - \alpha_g^{(\ell)} dP_{\mathrm{res}}^{(\ell)} .
\end{equation}
If one or more generators are predicted to violate limits, they are removed from the balancing set and clamped at the corresponding limit. The realized output is
\begin{equation}
    P_g^{(\ell+1)} =
    \begin{cases}
        P_g^{\mathrm{pred},(\ell)}, & \mathrm{if } P_g^{\min} < P_g^{\mathrm{pred},(\ell)} < P_g^{\max},\\
        P_g^{\min}, & \mathrm{if } P_g^{\mathrm{pred},(\ell)} \le P_g^{\min},\\
        P_g^{\max}, & \mathrm{if } P_g^{\mathrm{pred},(\ell)} \ge P_g^{\max}.
    \end{cases}
    \label{eq:generator_redistribution}
\end{equation}
Let $\mathcal{G}_s^{(\ell)}$ denote the set of generators clamped at their limits up to iteration $\ell$, and define the newly clamped set $\Delta\mathcal{G}_s^{(\ell+1)} = \mathcal{G}_s^{(\ell+1)} \setminus \mathcal{G}_s^{(\ell)}$. For each $s\in\Delta\mathcal{G}_s^{(\ell+1)}$, assign $\alpha_s^{(\ell+1)}=0$ and renormalize the remaining participation factors such that $\sum_{g\in\mathcal{G}_r^{(\ell+1)}} \alpha_g^{(\ell+1)} = 1$. The residual imbalance is then
\begin{equation}
    dP_{\mathrm{res}}^{(\ell+1)} = dP_{\mathrm{res}}^{(\ell)} + \sum_{s\in\Delta\mathcal{G}_s^{(\ell+1)}} \big(P_s^{(\ell+1)} - P_s\big).
    \label{eq:residual_allocation}
\end{equation}
This is repeated until the residual power is balanced ($dP_{\mathrm{res}}^{(\ell+1)}=0$) or no more participating generators remain ($\mathcal{G}_r^{(\ell+1)} = \varnothing$).

\vspace{0.25cm}
\subsubsection{Generator Reactive Power Response}
\label{sec:problem:response:reactive}

Active power balancing through participation factors uses system-wide frequency deviations as a shared feedback signal. Reactive power response lacks an analogous universal imbalance signal. Instead, voltage regulation is implemented through local excitation control, while wider-area coordination such as pilot-bus Automatic Voltage Control (AVC) is imposed by supervisory control based on zonal measurements and setpoint adjustments \cite{2016_avc}. We therefore adopt the standard AC-PF convention of representing all voltage-controlled buses as $PV$ buses, for which the bus voltage magnitude is held fixed while the generator reactive output satisfies $Q_g^{\min} \le Q_g \le Q_g^{\max}$ \cite{2012_overbye_book}. If a perturbation causes a generator’s reactive output to reach one of its limits, we assume saturation and fix $Q_g = Q_g^{\min}$ or $Q_g^{\max}$, thereby losing voltage regulation at the corresponding bus and converting it from $PV$ to $PQ$ operation \cite{1974_pf_pvpq, 2019_pv_pq_switching}. Once saturated, the generator is assumed to supply a fixed reactive injection, and the bus voltage is no longer directly controlled. We refer to this loss of voltage regulation as a $Q$-saturation regime change, discussed further in Section~\ref{sec:problem:response:regimes}.

\vspace{0.25cm}
\subsubsection{AC Network Response}
\label{sec:problem:response:network}
We now describe how the system state responds to perturbations in power injections and network topology. Local perturbations are modeled as continuous variations, while larger disturbances are treated as discrete events. Topological changes are represented through variations in network admittance elements $Y_{ij}$. Although physically implemented via discrete actions such as tap adjustments or switching of transformers and flexible AC transmission system (FACTS) devices, they are commonly approximated as continuous controls for sensitivity analysis. Generator and branch contingencies are permitted as discrete events.

Let $\mathcal{N}_p\subseteq\mathcal{N}$ denote the set of buses where active power injections are perturbed, and $\mathcal{N}_r\subseteq\mathcal{N}$ represent the buses where generators in $\mathcal{G}_r$ are connected. After redistribution in response to $dP_{\mathrm{inj}}=\sum_{i\in\mathcal{N}_p}dP_i^{\mathrm{(inj)}}$, the resulting active power perturbation at bus $i$ is
\begin{equation}
    dP_i =
    \begin{cases}
        dP_i^{\mathrm{(inj)}}, & \forall\, i \in \mathcal{N}_p, \\
        dP_i^{(l)}, & \forall\, i \in \mathcal{N}_r \setminus (\mathcal{N}_p \cap \mathcal{N}_r),\\
        0, & \forall\, i \notin (\mathcal{N}_p \cup \mathcal{N}_r).
    \end{cases}
    \label{eq:net_perturbation}
\end{equation}
The updated active power perturbation vector $d\mathbf{P}=[\,dP_i\,]^\top$ is used to define the new dispatch condition for the perturbed state, and is propagated through the AC network equations to obtain perturbed bus state for $\mathbf{P}$, $\mathbf{Q}$, $|\mathbf{V}|$, and~$\boldsymbol{\theta}$.

Let $\boldsymbol{\rho}=(\mathbf{u}, \mathbf{Y}_{\mathrm{bus}})$ collect the known parameters. The vector $\mathbf{u} = [\,\mathbf{P} \;\mathbf{Q}\,]^\top$ contains the specified active and reactive power injections, while $\mathbf{y}=[\,\boldsymbol{\theta} \;\mathbf{|V|}\,]^\top$ collects the corresponding voltage variables. In particular, $P$ is specified and $\theta$ is solved for at all non-slack buses $\mathcal{N}\setminus\{0\}$, while $Q$ is specified and $|V|$ is solved for at $PQ$ buses $\mathcal{N}_{PQ}$. Define bus power mismatches as 
\begin{equation}
    \mathbf{F}(\mathbf{y}; \boldsymbol{\rho})=
    \begin{bmatrix}
        \mathbf{P}(\mathbf{y}; \mathbf{Y}_{\mathrm{bus}}) - \mathbf{P} \\
        \mathbf{Q}(\mathbf{y}; \mathbf{Y}_{\mathrm{bus}}) - \mathbf{Q}
    \end{bmatrix} = \mathbf{G}(\mathbf{y}; \mathbf{Y}_{\mathrm{bus}}) - \mathbf{u}.
    \label{eq:mismatches}
\end{equation}
For a perturbation $d\boldsymbol{\rho}=(d\mathbf{u},d\mathbf{Y}_{\mathrm{bus}})$, the total differential of each component $F_k$ is
\begin{equation}
    dF_k =\sum_{m}\frac{\partial F_k}{\partial y_m}\,dy_m + 
          \sum_{m}\frac{\partial F_k}{\partial u_m}\,du_m +
          \sum_{i=1}^{N}\sum_{j=1}^{N}\frac{\partial F_k}{\partial Y_{ij}}\,dY_{ij}.
    \label{eq:mismatch_partials}
\end{equation}
Stacking all $k$ gives $d\mathbf{F}=[\,dF_k\,]^\top$. Compactly,
\begin{equation}
    d\mathbf{F} = \mathbf{F}_{\mathbf{y}}\,d\mathbf{y} +
                  \mathbf{F}_{\mathbf{u}}\,d\mathbf{u} +
                  \mathbf{F}_{\mathbf{Y}}\,:\,d\mathbf{Y}_{\mathrm{bus}},
    \label{eq:mismatch_change}
\end{equation}
with
\begin{equation}
    \mathbf{F}_{\mathbf{y}} = \frac{\partial \mathbf{F}}{\partial \mathbf{y}}, \quad
    \mathbf{F}_{\mathbf{u}} = \frac{\partial \mathbf{F}}{\partial \mathbf{u}}, \quad
    \mathbf{F}_{\mathbf{Y}} = \frac{\partial \mathbf{F}}{\partial \mathbf{Y}_{\mathrm{bus}}}.
    \label{eq:mismatch_matrices}
\end{equation}
We take the tensor contraction $\mathbf{F}_{\mathbf{Y}} : d\mathbf{Y}_{\mathrm{bus}}$ over a minimal basis spanning the admissible perturbation subspace of $\mathbf{Y}_{\mathrm{bus}}$, using only indices corresponding to perturbed branch admittances. Since $\mathbf{F}(\mathbf{y}; \boldsymbol{\rho})=\mathbf{G}(\mathbf{y}; \mathbf{Y}_{\mathrm{bus}}) - \mathbf{u}$, we then have
\begin{equation}
    \mathbf{F}_{\mathbf{y}} = \frac{\partial \mathbf{G}}{\partial \mathbf{y}} = \mathbf{J}, \quad
    \mathbf{F}_{\mathbf{u}} = -\mathbf{I}, \quad
    \mathbf{F}_{\mathbf{Y}} = \frac{\partial \mathbf{G}}{\partial \mathbf{Y}_{\mathrm{bus}}},
    \label{eq:mismatch_jacobian}
\end{equation}
with $\mathbf{J}$ being the Jacobian. Thus, in general form,
\begin{equation}
    d\mathbf{F} = \mathbf{J}(\mathbf{y};\boldsymbol{\rho})\,d\mathbf{y} - d\mathbf{u} + \mathbf{F}_{\mathbf{Y}}(\mathbf{y};\boldsymbol{\rho}) : d\mathbf{Y}_{\mathrm{bus}}.
    \label{eq:mismatch_change_nonlinear}
\end{equation}

At any equilibrium, there are no power imbalances. Then $\mathbf{F}(\mathbf{y};\boldsymbol{\rho}) = \mathbf{0}$, defines a smooth manifold of feasible operating points in the joint space of voltage variables and parameters. A perturbation $(d\mathbf{u}, d\mathbf{Y}_{\mathrm{bus}})$ induces displacements $(d\mathbf{y}, d\boldsymbol{\rho})$ that remain on this manifold if and only if the total differential $d\mathbf{F}(\mathbf{y};\boldsymbol{\rho})$ vanishes.

\vspace{0.25cm}
\paragraph{First-Order Approximation}
Consider a perturbation $\Delta \boldsymbol{\rho}$ that induces a change $\Delta\mathbf{y}$ such that the system settles to a new equilibrium. Since both the base and perturbed operating points satisfy the power-flow equations, we have
\begin{equation}
    \mathbf{F}(\mathbf{y}+\Delta\mathbf{y}; \boldsymbol{\rho}+\Delta \boldsymbol{\rho}) - \mathbf{F}(\mathbf{y};\boldsymbol{\rho}) = 0.
\end{equation}
Taylor expanding about $(\mathbf{y}_0,\boldsymbol{\rho}_0)$, we get
\begin{equation}
    \mathbf{J}_0\Delta\mathbf{y} - \Delta\mathbf{u} + \mathbf{F}_{\mathbf{Y}0}:\Delta\mathbf{Y}_{\mathrm{bus}} + \mathcal{O}(\|\Delta\mathbf{y}, \Delta\boldsymbol{\rho}\|^2) = 0,
    \label{eq:mismatch_firstorder_expansion}
\end{equation}
where $\mathbf{J}_0=\mathbf{J}(\mathbf{y}_0;\boldsymbol{\rho}_0)$, $\mathbf{F}_{\mathbf{Y},0}=\mathbf{F}_{\mathbf{Y}}(\mathbf{y}_0;\boldsymbol{\rho}_0)$, and $\mathcal{O}(\|\Delta\mathbf{y}, \Delta\boldsymbol{\rho}\|^2)$ represents all higher-order terms. To the first-order,
\begin{equation}
    \Delta\mathbf{y} \approx \mathbf{J}_0^{-1}\!\left(\Delta\mathbf{u} - \mathbf{F}_{\mathbf{Y},0} : \Delta\mathbf{Y}_{\mathrm{bus}}\right).
    \label{eq:voltage_change_firstorder}
\end{equation}
This first-order mapping describes the local sensitivity of bus voltages and angles to small injection and topology perturbations, assuming the Jacobian remains nonsingular and bus definitions remain unchanged.

\vspace{0.25cm}
\subsubsection{Monitored Quantities and First-Order Response}
\label{sec:problem:response:monitored}
Consider a monitored quantity, $x(\mathbf{y};\mathbf{Y}_{\mathrm{bus}})$, such as a bus state variable, generator output or branch flow. Under small perturbations in $\Delta \mathbf{u}$ and $\Delta \mathbf{Y}_{\mathrm{bus}}$, the first-order change in $x$ is 
\begin{equation}
    \Delta x = \frac{\partial x}{\partial \mathbf{y}}\bigg|_0\,\Delta\mathbf{y} +
        \frac{\partial x}{\partial \mathbf{Y}_{\mathrm{bus}}}\bigg|_0 : \Delta\mathbf{Y}_{\mathrm{bus}}.
    \label{eq:monitored_firstorder}
\end{equation}
Substituting \eqref{eq:voltage_change_firstorder} gives a practical decomposition into an injection term and a topology term:
\begin{align}
    \Delta x 
    & = \bigg(\frac{\partial x}{\partial \mathbf{y}}\bigg|_0 \mathbf{J}_0^{-1}\bigg) \Delta\mathbf{u} + \bigg(\frac{\partial x}{\partial \mathbf{Y}}\bigg|_0
        - \frac{\partial x}{\partial \mathbf{y}}\bigg|_0 \mathbf{J}_0^{-1}\mathbf{F}_{\mathbf{Y},0}\bigg) : \Delta\mathbf{Y}_{\mathrm{bus}} \nonumber \\
    & = \mathbf{L}_x \Delta\mathbf{u} + \mathbf{T}_x : \Delta\mathbf{Y}_{\mathrm{bus}}.
    \label{eq:monitored_firstorder_decomposed}
\end{align}
All derivatives and sensitivities are evaluated at the fixed operating point $(\mathbf{y}_0,\boldsymbol{\rho}_0)$ and define numerical linear operators once the base point is specified. While useful to characterize directional sensitivities, perturbed monitored quantities $x'$ can also be evaluated from the perturbed steady-state solution, $\mathbf{y}'$.

\vspace{0.25cm}
\subsubsection{Deterministic Regime Changes}
\label{sec:problem:response:regimes}
The network response above assumes generators control modes are fixed and the network topology remains well-defined. In practice, perturbations in injections or topology may trigger discrete regime changes associated with generator saturation or network islanding. These events introduce piecewise-smoothness to the feasible manifold, and can be treated sequentially. We consider regime changes in the following order:

\begin{enumerate}[itemsep=0pt, topsep=2pt]
    \item network admittance changes related to islanding,
    \item generator active power saturation, then finally
    \item generator reactive power saturation resulting in $PV\!\rightarrow\!PQ$ changes in bus definitions.
\end{enumerate}
$P$-saturation changes are detailed in Section~\ref{sec:problem:response:active}. We describe the remaining two classes of regime changes below.

\vspace{0.25cm}
\paragraph{$Y_{\mathrm{bus}}$ Regime Changes}
Discrete perturbations in branch admittances may induce changes in network connectivity when admittances are driven to zero. This effectively removes the connection and can electrically island buses. To avoid ill-conditioned or singular sensitivity matrices, the active system connected to the original slack bus must be reduced by removing these islanded components, which can then be handled separately based on modeling assumptions (e.g., ride-through control protocols or assumed blackout). Since generator control actions and $PV\!\rightarrow\!PQ$ transitions depend on connectivity, $Y_{\mathrm{bus}}$ regime changes must be resolved prior to any generator saturation sequencing.

\vspace{0.25cm}
\paragraph{$Q$-Saturation Regime Changes}
Following resolution of active power and network-admittance regime changes, reactive power limits are enforced through sequential $PV\!\rightarrow\!PQ$ transitions. Beginning from the solved base state, we compute the incremental change in voltages $\Delta \mathbf{y}$ in the current regime. From this, the first-order change in reactive power at each PV bus is obtained as
\begin{equation}
    \Delta Q_i = \frac{\partial Q_i}{\partial\mathbf{y}}\,\Delta \mathbf{y}
         + \frac{\partial Q_i}{\partial\mathbf{Y}_{\mathrm{bus}}}:\Delta\mathbf{Y}_{\mathrm{bus}}.
    \label{eq:reactive_sensitivity}
\end{equation}
The predicted reactive power $Q_i^{\mathrm{pred}}=Q_i+\Delta Q_i$ is then distributed to each generator $g$ at bus $i$ and compared to its limits $[Q_{g_i}^{\min},Q_{g_i}^{\max}]$. Generators for which $Q_{g_i}^{\mathrm{pred}}$ lies outside this range are flagged as in violation. If no violations are detected, the perturbation does not induce a regime change and the iteration ends. Otherwise, each violating generator is assigned a normalized event ratio
\begin{equation}
    t_{g_i}^{\mathrm{lim}} = \frac{Q_{g_i}^{\mathrm{lim}}-Q_{g_i}}{\Delta Q_{g_i}},
\end{equation}
where $t_{g_i}^{\mathrm{lim}}$ measures how far along the disturbance direction each output would change before its $Q$ limit is reached. The generator with the smallest positive $t_{g_i}^{\mathrm{lim}}$ is considered the first to reach a limit under the applied perturbation, and is clamped to the associated limit. If no more voltage-controlling generators exist at $i$, the bus is reclassified from $PV$ to $PQ$ and system equations are updated accordingly. The Jacobian and associated partial derivatives are reconstructed for the new regime, and the perturbation is remapped to obtain an updated state response. This process is repeated until no further $Q$-limit violations are detected.

\vspace{0.25cm}
\subsubsection{Discussion on Proximity to Stability Limits}
\label{sec:problem:response:stability}

Voltage stability phenomena are dynamical with nonlinear evolution, impacted by control interactions. In steady-state, voltage instability is associated with loss of regularity of the power-flow equations, which manifests as a singular Jacobian. Stability margins can be computed using CPF along prescribed loadability paths \cite{1992_cpf}. Implicit proximity measures can be derived from Jacobian-based indicators such as the smallest singular values, which provide a direction-independent measure of conditioning \cite{2007_voltage_stability}. Moreover, Jacobian singular value decomposition further insight by identifying combinations of power-balance equation imbalances for which small coordinated violations require large changes in bus voltage magnitudes and angles to restore equilibrium, often referred to as weak modes \cite{2007_voltage_stability}. These modes are intrinsic properties of the operating point and network structure and describe patterns of imbalance to which the system is vulnerable, independent of any specific disturbance direction being applied. 

Assessing proximity to stability limits along a specified $\Delta\boldsymbol{\rho}$ is conceptually closest to CPF methods. In principle, an analogous continuation procedure could be constructed by iteratively scaling $\Delta\boldsymbol{\rho}$, recomputing the power flow solution, and tracing the evolution of monitored quantities such as voltage magnitudes and reactive injections. As an initial fast screening, one could use the local linearized change $\Delta\mathbf{x}$ in response to $\Delta\boldsymbol{\rho}$ as a crude indicator for loss of regularity. If $\Delta x >\varepsilon_{\mathrm{reg}}$ or $\|\Delta\mathbf{x}\|>\varepsilon_{\mathrm{reg}}$ for some tolerance $\varepsilon_{\mathrm{reg}}$, the equilibrium response may be ill-conditioned along $\Delta\boldsymbol{\rho}$, and can trigger more detailed assessments.

\subsection{Off-Optimal Quasi-Dual Formulation}
\label{sec:problem:quasidual}
At an AC-OPF optimum, the KKT conditions imply that the cost gradient lies in the span of the gradients of the active constraints. The associated Lagrange multipliers
$\boldsymbol{\nu} = (\boldsymbol{\lambda}, \boldsymbol{\mu}, \boldsymbol{\sigma}^\pm, \boldsymbol{\tau}^\pm, \boldsymbol{\eta}^\pm, \boldsymbol{\pi}^\pm, \boldsymbol{\zeta})$
represent shadow prices associated with the power balance equations \eqref{eq:P_bal}--\eqref{eq:Q_bal} and active operational limits \eqref{eq:P_lims}--\eqref{eq:ang_ref}. These multipliers are well defined only at optimality, where the stationarity condition holds exactly.

Following a disturbance, the system settles to an equilibrium that is physically-feasible with respect to equality constraints \eqref{eq:P_bal} and \eqref{eq:Q_bal}, but possibly not operationally-feasible with respect to inequality constraints \eqref{eq:P_lims}--\eqref{eq:ang_ref}. Since no re-optimization occurred, the KKT conditions no longer hold and exact multipliers do not exist. Still, it is useful to retain multiplier-like quantities that describe how the cost gradient interacts with the local constraint manifold. This motivates an off-optimal quasi-dual construction, which extends the notion of shadow prices by projecting the cost gradient onto the span of active constraint gradients \cite{2006_quasiduals}. Let the primal decision vector be
\begin{equation}
    \mathbf{z} = (\mathbf{P}_{\mathcal{G}}, \mathbf{Q}_{\mathcal{G}}, \mathbf{V}_{\mathcal{N}}, \boldsymbol{\theta}_{\mathcal{N}}),
\end{equation}
with associated multipliers $\boldsymbol{\nu} \ge \mathbf{0}$. The pre-disturbance optimal equilibrium $(\mathbf{z}^\star, \boldsymbol{\nu}^\star)$ satisfies all KKT conditions. A small perturbation $(d\mathbf{u}, d\mathbf{Y}_{\mathrm{bus}})$ modifies injections and network parameters to
\begin{equation}
    \boldsymbol{\rho}' = \boldsymbol{\rho}^{\star} + d\boldsymbol{\rho} 
    = (\mathbf{u}^{\star} + d\mathbf{u}, \mathbf{Y}_{\mathrm{bus}}^{\star} + d\mathbf{Y}_{\mathrm{bus}}),
\end{equation}
resulting in a new physically-feasible off-optimal state
\begin{equation}
    \mathbf{z}' = \mathbf{z}^\star + d\mathbf{z}.
\end{equation}

\vspace{0.25cm}
\subsubsection{Constraint Manifold and Cost Gradient}
\label{sec:problem:quasidual:gradients}
Around any state $\mathbf{z}'$, the active constraints define a smooth manifold in the space of decision variables. The directions normal to that surface are given by the constraint gradients, and the directions along the surface are those that keep constraints satisfied to first order. Let the corresponding constraint-gradient matrix be
\begin{equation}
    \mathbf{C}(\mathbf{z}') =
    \begin{bmatrix}
        \nabla_{\mathbf{z}} \mathbf{h}(\mathbf{z}') \\
        \nabla_{\mathbf{z}} \mathbf{g}_{\mathcal{A}}(\mathbf{z}')
    \end{bmatrix},
    \label{eq:constraint_gradient}
\end{equation}
where $\mathbf{h}(\mathbf{z}')=\mathbf{0}$ collects the equality constraints, and $\mathbf{g}_{\mathcal{A}}(\mathbf{z}')\le \mathbf{0}$ collects the active and possibly violated inequality constraints at $\mathbf{z}'$. The cost gradient at $\mathbf{z}'$ is

\begin{equation}
    \nabla_{\mathbf{z}} f(\mathbf{z}') =
    \begin{bmatrix}
        \nabla_{\mathbf{P}_{\mathcal{G}}} f(\mathbf{P}_{\mathcal{G}}') \\
        \nabla_{\mathbf{Q}_{\mathcal{G}}} f(\mathbf{Q}_{\mathcal{G}}') \\
        \nabla_{\mathbf{|V|}_{\mathcal{N}}} f(\mathbf{|V|}_{\mathcal{N}}') \\
        \nabla_{\boldsymbol{\theta}_{\mathcal{N}}} f(\boldsymbol{\theta}_{\mathcal{N}}')
    \end{bmatrix} =
    \begin{bmatrix}
        \nabla_{\mathbf{P}_{\mathcal{G}}} f(\mathbf{P}_{\mathcal{G}}') \\
        \mathbf{0}_{\mathbf{Q}_{\mathcal{G}}} \\
        \mathbf{0}_{\mathbf{|V|_{\mathcal{N}}}} \\
        \mathbf{0}_{\boldsymbol{\theta}_{\mathcal{N}}}
    \end{bmatrix},
    \label{eq:cost_gradient}
\end{equation}
since only $\mathbf{P}_{\mathcal{G}}$ enters the objective \eqref{eq:primal_objective}. At $\mathbf{z}^\star$ the KKT stationarity condition can be written as
\begin{equation}
    \nabla_{\mathbf{z}} f(\mathbf{z}^\star) + \mathbf{C}(\mathbf{z}^\star)^\top \boldsymbol{\nu}^\star = 0, 
    \label{eq:KKT_stationarity}
\end{equation}
which means $\nabla_{\mathbf{z}} f(\mathbf{z}^\star)$ is a linear combination of constraint normals. At an AC feasible but suboptimal $\mathbf{z}'$
\begin{equation}
    \nabla_{\mathbf{z}} f(\mathbf{z}') + \mathbf{C}(\mathbf{z}')^\top \boldsymbol{\nu} \neq 0 \quad\mathrm{for}\;\mathrm{any}\;\boldsymbol{\nu},
    \label{eq:KKT_stationarity_invalid}
\end{equation}
and the nonzero residual is the component of the cost gradient that still points along the tangent directions, indicating possible cost reduction without violating constraints.

\vspace{0.25cm}
\subsubsection{Quasi-Duals from Least-Squares Projection}
\label{sec:problem:quasidual:definition}

We seek multiplier-like coefficients that make \eqref{eq:KKT_stationarity} hold as closely as possible at $\mathbf{z}'$. To do this, we perform a linear least-squares fit of the cost gradient by a combination of constraint normals. Define the quasi-dual vector at $\mathbf{z}'$ as
\begin{equation}
    \boldsymbol{\tilde{\nu}}(\mathbf{z}') = 
    \arg\min_{\boldsymbol{\nu}_{\mathcal A}}
    \big\| \nabla_{\mathbf{z}} f(\mathbf{z}') + \mathbf{C}(\mathbf{z}')^\top \boldsymbol{\nu} \big\|_2^2.
    \label{eq:quasidual}
\end{equation}
The quantity inside the norm is the \emph{stationarity residual}, which vanishes only at a true optimum.

This formulation does not impose sign restrictions on the components of $\boldsymbol{\tilde{\nu}}$. Instead, the quasi-duals are defined purely by the geometry of the cost gradient relative to the local constraint manifold. Minimizing \eqref{eq:quasidual} leads to the normal equations
\begin{equation}
    \mathbf{C}(\mathbf{z}')\,\mathbf{C}(\mathbf{z}')^\top \,\boldsymbol{\tilde{\nu}}(\mathbf{z}')
    = -\,\mathbf{C}(\mathbf{z}')\,\nabla_{\mathbf{z}} f(\mathbf{z}'),
    \label{eq:quasidual_normal_equations}
\end{equation}
and if $\mathbf{C}(\mathbf{z}')\,\mathbf{C}(\mathbf{z}')^\top$ is invertible, the closed-form solution is
\begin{equation}
    \boldsymbol{\tilde{\nu}}(\mathbf{z}') = 
        -\left(\mathbf{C}(\mathbf{z}')\,\mathbf{C}(\mathbf{z}')^\top\right)^{-1}
        \mathbf{C}(\mathbf{z}')\,\nabla_{\mathbf{z}} f(\mathbf{z}').
    \label{eq:quasidual_solved}
\end{equation}

Each component $\tilde{\nu}_k$ of $\boldsymbol{\tilde{\nu}}(\mathbf{z}')$ measures the signed projection of the unconstrained cost gradient onto the corresponding constraint normal. Its magnitude indicates how strongly that constraint shapes nearby cost-reducing directions. If $\tilde{\nu}_k>0$, the cost-reducing direction points into the constraint, so the constraint resists further cost reduction and acts as an economically active limit. If $\tilde{\nu}_k<0$, relaxing the constraint would reduce cost, indicating that it is not economically limiting at $\mathbf{z}'$, even if violated. Because quasi-duals depend on the local active set, they may vary discontinuously and are best interpreted as local geometric indicators.

\section{Deterministic Constraint Robustness}
\label{sec:deterministic}
We now characterize robustness with respect to operational constraints for deterministic perturbations. Let monitored quantity $x$ have bounds $L^{\min} \le x \le L^{\max}$. We define the associated constraint margin as
\begin{equation}
    \!\!M_x^{\mathrm{lim}} = s(L^{\mathrm{lim}} - x) \mathrm{, with}\; s=
    \begin{cases}
        +1 &\!\! \mathrm{, if}\; L^{\mathrm{lim}} \!=\! L^{\max}, \\
        -1 &\!\! \mathrm{, if}\; L^{\mathrm{lim}} \!=\! L^{\min},
    \end{cases}
\end{equation}
so that $M_x^{\mathrm{lim}} > 0$ indicates a respected limit for either bound. Except where needed, we drop ``lim'' superscripts for clarity, with $L$ and $M_x$ understood to either bound unless specified.

\subsection{Discrete Physical Robustness ($N\!-\!k$)}
\label{sec:deterministic:discrete}
For a $k^{\mathrm{th}}$-order discrete disturbance $\Delta\boldsymbol{\rho}$, the system is evaluated at the perturb state $\mathbf{z}'$ and the feasibility of all constraints is checked. If $M_x<0$ for any $x$, then a constraint is violated, and the perturbation is labeled as critical. A critical subset of violating quantities $\{x_{\mathrm{viol}}\}$ can be created for each critical perturbation for further consideration.

This corresponds directly to the classical $N\!-\!k$ screening procedure used in operational planning, where a system is ``$N\!-\!k$ secure'' if, for all contingencies in the screening set, a feasible solution exists that respects all limits \cite{2011_nmk_screening}. Whether as a post-solution screen or as part of an R-OPF formulation, it is also possible to allow for corrective actions between pre and post disturbance states. Here, we consider automated generator actions defined in Sections~\ref{sec:problem:response:active} and \ref{sec:problem:response:reactive}.

\subsection{Local Physical Robustness ($N\!+\!\delta_k $)}
\label{sec:deterministic:local}
Let $d\boldsymbol{\rho}$ be any $k^{\mathrm{th}}$-order perturbation direction, meaning any simultaneous variation of $k$ components of $d\boldsymbol{\rho}$. The directional sensitivity of $x$ with respect to $d\boldsymbol{\rho}$ is
\begin{equation}
    \frac{dx}{d\boldsymbol{\rho}} = 
    \frac{\partial x}{\partial \mathbf{y}}\frac{d\mathbf{y}}{d\boldsymbol{\rho}} + 
    \frac{\partial x}{\partial \boldsymbol{\rho}}.
    \label{eq:monitored_sensitivity}
\end{equation}
The directional change in the margin $M_x$ is
\begin{equation}
    \frac{dM_x}{d\boldsymbol{\rho}} = \frac{d\big(s(L - x)\big)}{d\boldsymbol{\rho}}
        = -s\frac{dx}{d\boldsymbol{\rho}}.
    \label{eq:margin_sensitivity}
\end{equation}
We define the \emph{sensitivity-normalized margin} as
\begin{equation}
    R_x = \frac{M_x}{dM_x/d\boldsymbol{\rho}}
         = \frac{-sM_x}{dx/d\boldsymbol{\rho}}, \quad M_x>0.
    \label{eq:sensitivity_normalized_margin}
\end{equation}

The index $R_x$ evaluated along $d\boldsymbol{\rho}$ is finite when $dM_x/d\boldsymbol{\rho}\neq 0$ and is infinite when the constraint is insensitive in first order. Its magnitude $|R_x|=M_x/|dM_x/d\boldsymbol{\rho}|$ represents the normalized distance to the limit in units of perturbation magnitude, while its sign indicates whether the margin shrinks or expands along $d\boldsymbol{\rho}$. Negative values identify shrinking margins, while positive values indicate expanding margins that are directionally robust even when $M_x$ is small.

We can use $R_x$ to create a local analog of $N\!-\!k$ security. If $-\varepsilon_R <R_x^{\mathrm{lim}}<0$ for a risk tolerance $\varepsilon_R>0$, we classify the pair $(x, \mathrm{lim})$ as locally vulnerable. We can thus say that an operating point is locally $N+\delta_k$ robust if, for every considered $k^{\mathrm{th}}$-order $d\boldsymbol{\rho}$, all $R_x$ satisfy $R_x \notin [-\varepsilon_R, 0)$.

\vspace{0.25cm}
\paragraph{Choice of Disturbance Basis}
For contingencies, the direction of $d\boldsymbol{\rho}$ is fixed by the loss of individual components. In contrast, the local formulation admits infinitely many perturbation directions within a neighborhood of $\boldsymbol{\rho}$, requiring the selection of a representative direction set for robustness screening. Lacking additional information, one may consider a \emph{cartesian-aligned basis} with equal signed $\pm$ variations to individual $P$, $Q$, $G$, or $B$ components, or a \emph{parameter-aligned basis} that scales perturbations along or orthogonal to $P+jQ$ or $G+jB$. In principle, any basis may be chosen. For example, local directions may be constructed to represent dominant modes of measurement or prediction error.

$R_x$ also depends on the normalization of $d\boldsymbol{\rho}$. Directional derivatives are defined with respect to unit-length directions, and are dimensionless. For screening purposes, however, it may be useful to retain a physical scaling of $\boldsymbol{\rho}$, such as a unit step as a 10~MW injection change or a 1~\% variation in line admittance. In this case, quantities such as $dM_x/d\boldsymbol{\rho}$ represent changes in margin per physically meaningful disturbance. Unit-normalized and physically-scaled $\boldsymbol{\rho}$ can thus be used to address different operational questions.

\subsection{Discrete Economic Robustness ($C\!-\!k$)}
\label{sec:deterministic:quasidual_discrete}
For a discrete $\Delta\boldsymbol{\rho}$, the system is evaluated at the perturbed operating point $\mathbf{z}'$ and the associated quasi-duals $\boldsymbol{\tilde{\nu}}(\mathbf{z}')$ are computed. Economic robustness is then assessed by comparing $\boldsymbol{\tilde{\nu}}(\mathbf{z}')$ against a prescribed cost-tolerance threshold $\varepsilon_{\nu}>0$, either element-wise or with a chosen norm. If any element or the chosen-norm exceeds $\varepsilon_{\nu}$, the perturbation is labeled as economically vulnerable.

This construction mirrors classical $N\!-\!k$ screening on the dual side. For a set of discrete $k$-order disturbances, the system is said to be $C\!-\!k$ economically robust if for all admissible $k^{\mathrm{th}}$-order perturbations, $\boldsymbol{\tilde{\nu}}(\mathbf{z}')$ or its elements are less than $\varepsilon_{\nu}$. $C\!-\!k$ robustness thus provides an economic analog to physical $N\!-\!k$ security, characterizing whether post-disturbance operating points remain economically well-conditioned.

\subsection{Local Economic Robustness ($C\!+\!\delta_k$)}
\label{sec:deterministic:quasidual_local}
Section~\ref{sec:deterministic:local} introduced the sensitivity-normalized margin $R_x$ which measures, to first order, how far a constraint remains feasible under an infinitesimal disturbance direction~$d\boldsymbol{\rho}$. A natural question is whether an analogous quantity can be defined on the dual side.  In principle, one may consider a dual margin $M_x^{\nu} = \nu_x^{\max} - \nu_x$, and the associated sensitivity-normalized margin
\begin{equation}
    R_x^{\nu} = \frac{M_x^{\nu}}{d\nu_x/d\boldsymbol{\rho}}, \quad M_x^{\boldsymbol{\nu}}\ge0,
\end{equation}
which would represent the first-order change in $\nu_x$ associated with $x$ toward a shadow-price limit $\nu_x^{\max}$. However, two issues arise. First, many markets do not consider explicit price caps, making $\nu_x^{\max}$ undefined. Second, computing $d\nu_x/d\boldsymbol{\rho}$ requires differentiating the quasi-dual system \eqref{eq:quasidual_normal_equations}, with respect to the disturbance direction~$d\boldsymbol{\rho}$. This introduces derivatives of both the active-constraint matrix $\mathbf{C}(\mathbf{z}')$ and the cost gradient $\nabla_{\mathbf{z}} f(\mathbf{z}')$, producing a second-order system that is comparatively more expensive to evaluate.

A simple alternative is to characterize how strongly each active constraint contributes to the change in marginal cost along~$d\boldsymbol{\rho}$. Using the standard KKT decomposition
\begin{equation}
    \nabla_{\mathbf{z}} f(\mathbf{z}) \approx \mathbf{C}^\top \boldsymbol{\nu}.
\end{equation}
The directional cost sensitivity then satisfies
\begin{equation}
    \frac{df}{d\boldsymbol{\rho}} 
        \approx \boldsymbol{\nu}^\top \mathbf{C}\,\frac{d\mathbf{z}}{d\boldsymbol{\rho}} 
        = \sum_{a\in\mathcal{A}} \nu_a \frac{dg_a}{d\boldsymbol{\rho}},
\end{equation}
where $dg_a/d\boldsymbol{\rho}$ is the directional change of the constraint function induced by the state response. Equivalently, this directional change satisfies $dg_x/d\boldsymbol{\rho} = -\,dM_x/d\boldsymbol{\rho}$. 

We define the \emph{shadow-price volatility index} (SPVI) for a constrained monitored quantity as
\begin{equation}
    \varsigma_x = \nu_x \frac{dg_x}{d\boldsymbol{\rho}} 
           = - \nu_x \frac{dM_x}{d\boldsymbol{\rho}},
\end{equation}
which measures how strongly the binding constraint on $x$ contributes to the local cost sensitivity along~$d\boldsymbol{\rho}$. To turn this into a screening tool, we introduce a volatility tolerance $\varepsilon_{\mathrm{SPVI}}>0$ and classify each active constraint $k$ as locally cost-robust along~$\rho$ if $|\varsigma_x(\rho)| < \varepsilon_{\mathrm{SPVI}}$, and cost-vulnerable otherwise. Similar to $N\!+\!\delta_k$ physical robustness screening, a system is said to be $C\!+\!\delta_k$ economically robust with respect to a family of $k^{\mathrm{th}}$-order disturbance directions if every active constraint is cost-robust. 

\section{Probabilistic Constraint Robustness}
\label{sec:probabilistic}
The screening tools in Section~\ref{sec:deterministic} assess how operating margins change in response to a specified perturbation or direction. In practice, forecast errors and operational variability are often random. It is thus valuable to characterize not only the distance to constraints but also the probabilities of constraint violation. Here we build on PLF foundations to extend the deterministic framework to random perturbations, using the first-order mappings derived in Section~\ref{sec:problem}. We first consider perturbations within a fixed network and control regime, such that the linearized response is well defined. We then address regime changes in Section~\ref{sec:probabilistic:regimes}.

\subsection{Linear Primal Uncertainty Mapping}
\label{sec:probabilistic:linear}

In the deterministic setting, perturbations in injections $\Delta\mathbf{u}$ and network parameters $\Delta\mathbf{Y}_{\mathrm{bus}}$ appear as separate objects. In the probabilistic setting, however, uncertain injections and topology parameters in general act as a \emph{joint} disturbance with possible covariance. We therefore collect all perturbations into a single real-valued stacked disturbance vector,
\begin{equation}
    \Delta\boldsymbol{\rho} =
    \begin{bmatrix}
        \Delta\mathbf{u} \\
        \mathrm{vec}(\Delta\mathbf{G}_{\mathrm{bus}}) \\
        \mathrm{vec}(\Delta\mathbf{B}_{\mathrm{bus}})
    \end{bmatrix},
    \label{eq:stacked_perturbations_real}
\end{equation}
where $\Delta\mathbf{Y}_{\mathrm{bus}} = \Delta\mathbf{G}_{\mathrm{bus}} + j\,\Delta\mathbf{B}_{\mathrm{bus}}$. As in Section~\ref{sec:problem:response:network}, uncertainty in $\mathbf{Y}_{\text{bus}}$ is restricted to a minimal basis that spans the support of admissible network uncertainty such that 
\begin{equation}
    \Delta\mathbf{Y}_{\text{bus}} = \sum_{k=1}^{K} \delta y_k \,\mathbf{E}_k,
    \label{eq:Ybus_basis_global}
\end{equation}
where $\{\mathbf{E}_k\}_{k=1}^{K}$ identify the nonzero perturbation directions permitted by the uncertainty model, and $\boldsymbol{\delta}_Y = [\delta y_1,\ldots,\delta y_K]^\top$ is the associated vector of random coefficients. This basis is not assumed orthogonal or unique, and serves only to identify the support of admissible network perturbations. Separating real and imaginary components yields
\begin{equation}
    \mathrm{vec}(\Delta\mathbf{G}_{\text{bus}}) = \mathbf{E}_G \boldsymbol{\delta}_G,
    \quad
    \mathrm{vec}(\Delta\mathbf{B}_{\text{bus}}) = \mathbf{E}_B \boldsymbol{\delta}_B,
    \label{eq:GB_embedding}
\end{equation}
where $\mathbf{E}_G$ and $\mathbf{E}_B$ are fixed embedding matrices, and $\boldsymbol{\delta}_G,\boldsymbol{\delta}_B$ collect the real-valued network uncertainty parameters.

For a scalar monitored quantity $x$, the linearized response can then be written as
\begin{align}
    \Delta x = \mathbf{L}_x\,\Delta\mathbf{u}
        &+ \mathrm{vec}(\mathbf{T}_{x,G})^\top \mathrm{vec}(\Delta\mathbf{G}_{\mathrm{bus}}) \nonumber \\
        &+ \mathrm{vec}(\mathbf{T}_{x,B})^\top \mathrm{vec}(\Delta\mathbf{B}_{\mathrm{bus}}), \nonumber \\
        = \mathbf{L}_x\,\Delta\mathbf{u}
        &+ \mathbf{S}_{x,G}\,\boldsymbol{\delta}_G
        + \mathbf{S}_{x,B}\,\boldsymbol{\delta}_B,
    \label{eq:monitored_change_real}
\end{align}
where $\mathbf{T}_{x,G} = \partial x / \partial \mathbf{G}_{\text{bus}}$ and $\mathbf{T}_{x,B} = \partial x / \partial \mathbf{B}_{\text{bus}}$ are evaluated at the operating point associated with the current regime, and $\mathbf{S}_{x,G} = \mathrm{vec}(\mathbf{T}_{x,G})^\top \mathbf{E}_G$ and $\mathbf{S}_{x,B} = \mathrm{vec}(\mathbf{T}_{x,B})^\top \mathbf{E}_B$. 

Collecting terms yields the compact representation
\begin{equation}
    \Delta x = \mathbf{S}_x \,\Delta\boldsymbol{\rho},
    \label{eq:monitored_change_stacked}
\end{equation}
with $\mathbf{S}_x = \big[ \mathbf{L}_x \;\;  \mathrm{vec}(\mathbf{T}_{x,G})^\top \;\;  \mathrm{vec}(\mathbf{T}_{x,B})^\top \big]$. 

If $\Delta\boldsymbol{\rho}$ has finite first and second moments,
\begin{equation}
    \mathbb{E}[\Delta\boldsymbol{\rho}] = \boldsymbol{\mu}_\rho, \qquad
    \mathrm{Cov}(\Delta\boldsymbol{\rho}) = \boldsymbol{\Sigma}_\rho,
    \label{eq:pert_mean_covar}
\end{equation}
then by linearity in \eqref{eq:monitored_change_stacked}, the induced moments of $\Delta x$ are
\begin{align}
    \mu_x &= \mathbb{E}[\Delta x] = \mathbf{S}_x \boldsymbol{\mu}_\rho, \label{eq:monitored_mean} \\
    \sigma_x^2 &= \mathrm{Var}(\Delta x) = \mathbf{S}_x \boldsymbol{\Sigma}_\rho \mathbf{S}_x^\top. \label{eq:monitored_var}
\end{align}

\subsection{Moment-Based Constraint Margins}
\label{sec:probabilistic:moment}

Consider $x$ with limits $L_{\min} \le x \le L_{\max}$, base value $x^\star$, and margin $M_x^\star \ge 0$. Under a random perturbation $\Delta\boldsymbol{\rho}$, the linearized response $\Delta x = \mathbf{S}_x \Delta\boldsymbol{\rho}$ results in a constraint violation whenever
\begin{equation}
    s\,\Delta x \ge M_x^\star.
    \label{eq:violation_event}
\end{equation}
For a binding constraint with $M_x^\star=0$, $x$ is released when 
\begin{equation}
    s\,\Delta x < 0.
\end{equation}
To support robustness screening, we define two moment-based margins. First, we define the \emph{$\sigma$-normalized margin} as 
\begin{equation}
    R_{\sigma,x} = \frac{M_x^\star - s\,\mu_x}{\sigma_x}, 
    \label{eq:var_distance}
\end{equation}
which measures the distance between the limit and the mean drift of $x$, in standard deviations of $\Delta x$. We next define the one-sided upper-tail $\alpha$-quantile \cite{2003_utoronto_stat_text} of $s\,\Delta x$ by $q_{\alpha,x}$ such that
\begin{equation}
    \Pr\!\left(s\,\Delta x \le q_{\alpha,x}\right) = \alpha
    \;\Longleftrightarrow\;
    \Pr\!\left(s\,\Delta x \ge q_{\alpha,x}\right) = 1-\alpha.
    \label{eq:q_alpha_signed}
\end{equation}
We define the \emph{$\alpha$-confidence margin} as
\begin{equation}
    M_{\alpha,x} = M_x^\star - q_{\alpha,x},
    \label{eq:alpha_conf_margin_def}
\end{equation}
so that $M_{\alpha,x} > 0$ implies $\Pr(s\,\Delta x \le M_x^\star)\ge \alpha$. Using the first two moments, we have $q_{\alpha,x} = s\,\mu_x + \xi_\alpha\,\sigma_x$. Then 
\begin{equation}
    M_{\alpha,x} = M_x^\star - s\,\mu_x - \xi_\alpha\,\sigma_x.
    \label{eq:quantile_margin}
\end{equation}

\vspace{0.25cm}
\paragraph{Choice of One-sided Quantile}
The formulation above applies for any uncertain disturbance with finite first and second moments. When only these moments are known, Cantelli’s inequality \cite{1952_cantelli_in_review} provides a distribution-free upper bound on the one-sided tail of $s\,\Delta x$. For any $\alpha\in(0,1)$,
\begin{equation}
    \Pr\!\left( s\,\Delta x \ge s\,\mu_x + \sqrt{\frac{\alpha}{1-\alpha}}\,\sigma_x \right)
        \;\le\; 1-\alpha.
    \label{eq:cantelli_alpha}
\end{equation}
Equivalently, with probability at least $\alpha$,
\begin{equation}
    s\,\Delta x \;\le\; s\,\mu_x + \xi_\alpha^{(C)}\,\sigma_x, \quad
    \xi_\alpha^{(C)} = \sqrt{\frac{\alpha}{1-\alpha}}.
\end{equation}

When the distribution of $\Delta x$  is known, the one-sided score $\xi_\alpha$ may be evaluated directly from the corresponding cumulative distribution function (CDF). If $\Delta\boldsymbol{\rho}$ is modeled as jointly normal, then $s \Delta x$ is normally distributed and
\begin{equation}
    \xi_\alpha^{(N)} = \Phi^{-1}(\alpha),
    \label{eq:gaussian_score}
\end{equation}
where $\Phi(\cdot)$ is the standard normal CDF. For example, $\xi_{0.95}^{(N)}=1.645$, recovers the normal 95\% one-sided quantile.

\subsection{Distribution-Free Bounds on Constraint Violation and Release Probabilities}
\label{sec:probabilistic:cantelli}
Often, not enough data exist to quantify distributional uncertainties reliably. When only the first and second moments are available, we can nevertheless obtain conservative bounds on violation and release probabilities. 

\vspace{0.25cm}
\subsubsection{Upper Bound on Violation Probabilities}
When $s\mu_x < M_x^\star$, we can obtain an upper bound on the probability of constraint violation. Applying Cantelli's inequality gives
\begin{equation}
    \Pr\big(s \Delta x \!\ge\! M_x^\star\big) \le \frac{\sigma_x^2}{\sigma_x^2 \!+\! (M_x^\star \!-\! s\,\mu_x)^2}
    = \frac{1}{1 \!+\! R_{\sigma,x}^2}.
    \label{eq:cantelli_bound_monitored}
\end{equation}
The resulting bound remains valid for any underlying distribution with the same moments, whether non-Gaussian, asymmetric, or heavy-tailed. The condition $s\,\mu_x < M_x^\star$ indicates that the mean drift remains inside the feasible region. If $s\,\mu_x \ge M_x^\star$, \eqref{eq:cantelli_bound_monitored} reduces to the trivial bound $\Pr(s\,\Delta x \ge M_x^\star)\le 1$, since it must also hold when $\sigma_x=0$. 

\vspace{0.25cm}
\subsubsection{Lower Bound on Release Probabilities}
For binding constraints when $s\,\mu_x < 0$, we can calculate a lower bound on the release probability. Applying Cantelli's inequality gives
\begin{align}
    \Pr(s\,\Delta x < 0) 
        &= 1- \Pr(s\,\Delta x \ge 0) \label{eq:general_release} \\
        &\ge 1- \frac{1}{1 + R_{\sigma,x}^2} = \frac{R_{\sigma,x}^2}{1 + R_{\sigma,x}^2},
\end{align}
with $M_x^{\star}=0$ and thus $R_{\sigma,x}=- s\,\mu_x/\sigma_x$. If $s\,\mu_x \ge 0$, we get the trivial bound $\Pr(s\,\Delta x < 0)\ge0$.

\subsection{Constraint Violation and Release Probabilities for Disturbance Distributions}
\label{sec:probabilistic:exact}
When the joint distribution of the uncertain disturbance vector $\Delta\boldsymbol{\rho}$ is defined, the affine map \eqref{eq:monitored_change_stacked} induces well-defined distributions for $\Delta x$. This enables exact evaluations of constraint violation and release probabilities.

\vspace{0.25cm}
\subsubsection{Violation Probabilities for Gaussian Uncertainty}
A common assumption in PLF and OPF studies is that $\Delta\boldsymbol{\rho}$ follows a multivariate Gaussian distribution \cite{2017_plf_review}. Under this assumption, the affine response $\Delta x$ is itself Gaussian, with mean $\mu_x$ and variance $\sigma_x^2$ as in \eqref{eq:monitored_mean} and \eqref{eq:monitored_var}. Under Gaussian uncertainty, the violation probability is
\begin{align}
    \Pr(s\,\Delta x \ge M_x^\star) 
        &= \Pr\!\left(\frac{s\,\Delta x - s\,\mu_x}{\sigma_x}
            \ge \frac{M_x^\star - s\,\mu_x}{\sigma_x}\right) \nonumber \\
        &= 1 - \Phi(R_{\sigma,x}).
    \label{eq:violation_gaussian}
\end{align}

\vspace{0.25cm}
\subsubsection{Release Probabilities for Gaussian Uncertainty}
For binding constraints, the release probability is
\begin{equation}
    \Pr(s\,\Delta x \!<\! 0) = 
        \Pr\!\left(\!\frac{s\,\Delta x \!-\! s\,\mu_x}{\sigma_x} \!<\! 
            \frac{-s\,\mu_x}{\sigma_x}\!\right) \!= \Phi(R_{\sigma,x}).
    \label{eq:release_gaussian}
\end{equation}

\subsection{Linear Quasi-Dual Uncertainty Mapping}
\label{sec:probabilistic:quasidual}
In Section~\ref{sec:problem:quasidual}, deterministic quasi-duals are constructed using an already calculated $\mathbf{z}'$ by projecting the cost gradient onto the span of constraint normals active at that point. In the probabilistic setting, however, $\mathbf{z}'$ is random, and both the active constraint set and the local constraint geometry vary across realizations. As a result, $\boldsymbol{\tilde{\nu}}(\mathbf{z}')$ is not directly amenable to linearization with respect to $\mathbf{z}'$ over the full uncertainty support. To obtain a tractable approximation, we instead propagate uncertainty from $\Delta\boldsymbol{\rho}$ into the quasi-dual space using a linearization evaluated at a representative operating point, and restrict the resulting quasi-dual variations to domains where the corresponding constraints are applicable. For compactness, define the projected cost gradient
\begin{equation}
    \mathbf{q}(\mathbf{z}) = \mathbf{C}(\mathbf{z})\,\nabla_{\mathbf{z}} f(\mathbf{z}),
    \label{eq:quasidual_q_prob}
\end{equation}
and the normal matrix
\begin{equation}
    \mathbf{H}(\mathbf{z}) = \mathbf{C}(\mathbf{z})\mathbf{C}(\mathbf{z})^\top,
    \label{eq:quasidual_H_prob}
\end{equation}
so that, within a region with a fixed set of active constraints, the quasi-duals satisfy
\begin{equation}
    \boldsymbol{\tilde{\nu}}(\mathbf{z}) = -\,\mathbf{H}(\mathbf{z})^{-1}\mathbf{q}(\mathbf{z}).
    \label{eq:quasidual_map_prob}
\end{equation}
Let
\begin{equation}
    \bar{\mathbf{z}} = \mathbb{E}[\mathbf{z}']
    \approx \mathbf{z}^\star + \mathbb{E}[\Delta\mathbf{z}]
    \label{eq:zbar_prob}
\end{equation}
denote the mean perturbed state obtained from the linear primal uncertainty mapping. While the active set is fixed, the quasi-dual mapping \eqref{eq:quasidual_map_prob} is smooth. The differential is
\begin{equation}
    d\boldsymbol{\tilde{\nu}}
        = -\,d\!\left(\mathbf{H}^{-1}\right)\mathbf{q} - \mathbf{H}^{-1}\,d\mathbf{q}.
    \label{eq:dnu_step1}
\end{equation}
Using the identity $d\!\left(\mathbf{H}^{-1}\right) = -\,\mathbf{H}^{-1}(d\mathbf{H})\,\mathbf{H}^{-1}$ and substituting $\mathbf{q}=-\mathbf{H}\boldsymbol{\tilde{\nu}}$, we obtain
\begin{equation}
    d\boldsymbol{\tilde{\nu}}
    = -\,\mathbf{H}^{-1} d\mathbf{q} - \mathbf{H}^{-1}(d\mathbf{H})\,\boldsymbol{\tilde{\nu}}.
    \label{eq:dnu_step2}
\end{equation}
Next, differentiating \eqref{eq:quasidual_q_prob}--\eqref{eq:quasidual_H_prob} gives
\begin{align}
    d\mathbf{H}
    &= d(\mathbf{C}\mathbf{C}^\top)
    = (d\mathbf{C})\,\mathbf{C}^\top + \mathbf{C}\,(d\mathbf{C})^\top,
    \label{eq:dH_expand} \\
    d\mathbf{q}
    &= d(\mathbf{C}\nabla_{\mathbf{z}} f)
    = (d\mathbf{C})\,\nabla_{\mathbf{z}} f + \mathbf{C}\,d(\nabla_{\mathbf{z}} f).
    \label{eq:dq_expand}
\end{align}
Both $d\mathbf{C}$ and $d(\nabla_{\mathbf{z}} f)$ depend linearly on $d\mathbf{z}$ through second derivatives (constraint and cost Hessians). Writing the differential columnwise, for each state component $z_k$,
\begin{equation}
    \frac{\partial \boldsymbol{\nu}}{\partial z_k}
    = -\,\mathbf{H}^{-1} \Bigg[
        \frac{\partial\mathbf{q}}{\partial z_k} 
        + \frac{\partial\mathbf{H}}{\partial z_k}\,\boldsymbol{\nu} \Bigg],
    \label{eq:dnu_column_compact}
\end{equation}
with
\begin{align}
    \frac{\partial\mathbf{H}}{\partial z_k}
        &= \frac{\partial\mathbf{C}}{\partial z_k}\,\mathbf{C}^\top
            + \mathbf{C}\left(\frac{\partial\mathbf{C}}{\partial z_k}\right)^\top,
    \label{eq:dH_dzk} \\
    \frac{\partial\mathbf{q}}{\partial z_k}
        &= \frac{\partial\mathbf{C}}{\partial z_k}\,\nabla_{\mathbf{z}} f
            + \mathbf{C}\,\frac{\partial^2 f}{\partial \mathbf{z}\,\partial z_k}.
    \label{eq:dq_dzk}
\end{align}
Collecting \eqref{eq:dnu_column_compact} over all $k$ gives the compact form
\begin{equation}
    d\boldsymbol{\tilde{\nu}} = \mathbf{K}_{\boldsymbol{\tilde{\nu}}}^{(\bar{\mathbf{z}})}\,d\mathbf{z},
    \label{eq:Knu_def}
\end{equation}
where $\mathbf{K}_{\boldsymbol{\nu}}^{(\bar{\mathbf{z}})}$ is evaluated at $\mathbf{z}=\bar{\mathbf{z}}$ and uses the active set $\mathcal{A}(\bar{\mathbf{z}})$. Using the linearized state response $\Delta\mathbf{z} = \mathbf{S}_z\,\Delta\boldsymbol{\rho}$, we get
\begin{equation}
    \Delta\boldsymbol{\tilde{\nu}}
        \approx \mathbf{K}_{\boldsymbol{\tilde{\nu}}}\,\Delta\boldsymbol{\rho}, \quad
    \mathbf{K}_{\boldsymbol{\tilde{\nu}}} 
        = \mathbf{K}_{\boldsymbol{\tilde{\nu}}}^{(\bar{\mathbf{z}})}\,\mathbf{S}_z,
    \label{eq:quasidual_affine_prob}
\end{equation}
which describes the linear propagation of uncertainty into the quasi-dual space under the assumption of fixed constraint geometry. Under this approximation, the first and second moments of $\Delta\boldsymbol{\nu}$ follow directly from those of $\Delta\boldsymbol{\rho}$:
\begin{align}
    \mathbb{E}[\Delta\boldsymbol{\tilde{\nu}}]
        &= \mathbf{K}_{\boldsymbol{\tilde{\nu}}}\,\boldsymbol{\mu}_\rho,
    \label{eq:quasidual_mean_prob} \\
    \mathrm{Cov}(\Delta\boldsymbol{\tilde{\nu}})
        &= \mathbf{K}_{\boldsymbol{\tilde{\nu}}}\,\boldsymbol{\Sigma}_\rho\,
           \mathbf{K}_{\boldsymbol{\tilde{\nu}}}^\top.
    \label{eq:quasidual_cov_prob}
\end{align}

The linear mapping \eqref{eq:quasidual_affine_prob} is valid only within regions where the active set remains unchanged, the regime remains unchanged, and the first-order approximation holds. We do not attempt to quantify the size of this region or the probability of active-set changes here. 

\vspace{0.25cm}
\subsubsection{Inequality Quasi-Dual Domain Restriction}
Quasi-duals associated with inequality constraints are physically meaningful only when the corresponding constraints are binding or violated. We then restrict each component of $\Delta\boldsymbol{\tilde{\nu}}$ to an admissible domain defined by the corresponding constraint margin. Define the admissible domain of $\tilde{\nu}_x$ as 
\begin{equation}
    \mathcal{D}_x := \{\,\Delta\boldsymbol{\rho} : s\,\Delta x \ge M_x^\star\,\}.
    \label{eq:admissible_domain}
\end{equation}
Since $\Delta x = \mathbf{S}_x\,\Delta\boldsymbol{\rho}$, the domain $\mathcal{D}_x$ corresponds to a half-space in $\Delta\boldsymbol{\rho}$. The truncated quasi-dual variation associated with constraint $x$ is then defined as
\begin{equation}
    \Delta \bar \nu_x =
    \begin{cases}
        \Delta\tilde{\nu}_x, & \text{if } \Delta\boldsymbol{\rho} \in \mathcal{D}_x, \\[0.3em]
        0, & \text{otherwise}.
    \end{cases}
    \label{eq:nu_truncated_prob}
\end{equation}
Equivalently, truncation occurs at the hyperplane $s\,\Delta x = M_x^\star$, which serves as the activation threshold for the corresponding quasi-dual component. 

$\mathcal{D}_x$ is evaluated using the same linear response $\Delta x$. When the distribution of $\Delta x$ is known, the probability of activation $\Pr(\mathcal{D}_x)$ can be computed directly. When only first and second moments are available, conservative bounds on $\Pr(\mathcal{D}_x)$ can be obtained using distribution-free inequalities, as discussed in Section~\ref{sec:probabilistic:cantelli}.

\subsection{Probabilistic Regime Changes}
\label{sec:probabilistic:regimes}
Sections~\ref{sec:problem:response:active}--\ref{sec:problem:response:regimes} resolve generator and network discontinuities through regime changes associated with active power limits, network connectivity, and reactive power limits. These regime classes differ in how a unique ordering of events can be defined. In the deterministic setting, all detected active power saturation events are treated as simultaneous, and higher-order network connectivity changes are likewise modeled as a single event by definition of $N\!-\!k$ contingencies. Reactive power saturation admits a natural ordering using directional voltage sensitivities and predicted overshoots to define a unique sequence of $Q$-limit activations as generators transition from $PV$ to $PQ$ operation.

For uncertain disturbances, these deterministic sequences are only a subset of the possible regime evolutions. Different orderings and combinations of regime changes may occur with non-negligible probability, based on how probability masses interact with operational limits and connectivity thresholds.

Ignoring event order, each generator with potential active- or reactive power saturation admits three possible states (lower limit, interior, or upper limit), while each candidate disconnection admits two states (connected or disconnected). The number of regime patterns is thus bounded by
\begin{equation}
    N_{\mathrm{pattern}} \le 3^{N_P+N_Q}\,2^{N_Y},
\end{equation}
where $N_P$ and $N_Q$ denote the numbers of generators with possible $P$- and $Q$-saturation events, and $N_Y$ the number of branches with possible disconnection events.

If event order is also considered, the combinatorial space grows substantially. Allowing any subset of the $N_P+N_Q+N_Y$ candidate events to occur in arbitrary sequence gives at most
\begin{equation}
    N_{\mathrm{sequence}} \le \sum_{k=0}^{N_P+N_Q+N_Y}
    \frac{(N_P+N_Q+N_Y)!}{(N_P+N_Q+N_Y-k)!}
\end{equation}
sequences, which grows superexponentially with system size \cite{2002_stat_infr_book}. Exhaustive enumeration quickly becomes intractable.

To address this, we introduce two order-reduction methods. For distributional uncertainties, we construct a sequentially-pruned regime tree that enumerates physically consistent and probabilistically relevant regime evolutions while preserving exact probability mass. When only moments are known or used, we construct an $\alpha$-stressed regime for evaluating conservative bounds on violation and release probabilities.

\vspace{0.25cm}
\subsubsection{Sequential-Pruning Mixture Representation Under Known Disturbance Distributions}
\label{sec:probabilistic:mixture}

Assume a known distribution for $\Delta\boldsymbol{\rho}$. We construct a regime tree where each tree node $n$ represents a regime $\mathcal{R}_n$ defined by a conditioning region in disturbance space, together with a regime-local conditional disturbance $\Delta\boldsymbol{\rho}^{(n)} \sim \Delta\boldsymbol{\rho}\mid\mathcal{R}_n$ and an associated linear response model $\mathbf{S}^{(n)}$. An edge from parent node $m$ to child node $n$ corresponds to a single regime-change event and carries conditional probability $p_{n\mid m} = \Pr(\mathcal{R}_n \mid \mathcal{R}_m)$, so that each node has unconditional probability mass $\gamma_n = \gamma_m\,p_{n\mid m}$, with $\gamma_0=1$ for the root node.

Tree construction proceeds sequentially by regime class. At each non-terminal node $m$, a single candidate regime-change event is selected and tested, inducing a binary split of $\mathcal{R}_m$ into two complementary outcomes. If both resulting child probabilities exceed a prescribed pruning threshold $\varepsilon_p$, both children are added as nodes. If only one outcome exceeds $\varepsilon_p$, the parent node is updated in place to represent the surviving conditional regime, and the complementary low-probability outcome is collapsed into a probability atom to the corresponding limit event. This path-compression strategy preserves probability mass while minimizing tree growth. A node is terminal for a given regime class when no further candidate events or controllable elements remain.

\begin{figure}
    \centering
    \includegraphics[width=\linewidth]{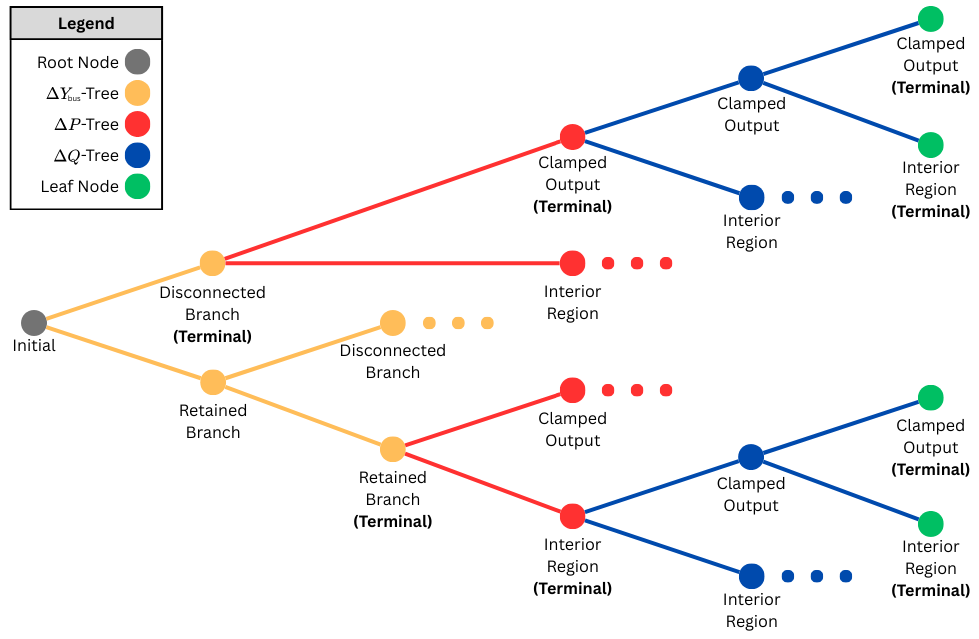}
    \caption{Hierarchical composition of the sequentially-pruned regime tree, illustrating the ordered attachment of sub-trees.}
    \label{fig:tree}
\end{figure}

The full regime tree is assembled by attaching sub-trees in the same regime-class order as in Section~\ref{sec:problem:response:regimes}. A $\Delta\mathbf{Y}_{\mathrm{bus}}$ sub-tree is constructed first to resolve connectivity changes. To each of its terminal nodes, a $\Delta\mathbf{P}$ sub-tree is attached to represent active power saturation. Finally, $\Delta\mathbf{Q}$ sub-trees are attached to capture reactive power saturation and $PV\!\rightarrow PQ$ transitions. Each leaf of the resulting composite tree corresponds to a realized sequence of regime changes and accumulates the conditional probabilities along its path from the root. This structure is illustrated in Figure~\ref{fig:tree}.

Each node implicitly represents a conditioning region $\mathcal{R}_n$ defined by the intersection of branching events along its path. The resulting tree does not enumerate all possible regime orderings, nor does it attempt to represent exact regime-local probability densities. It provides an accounting of physically consistent regime evolutions whose conditional probability exceeds the prescribed tolerance.

\vspace{0.25cm}
\paragraph{$\Delta \mathbf{Y}_{\mathrm{bus}}$-Tree Construction}
Let $\Delta\mathbf{Y}_{\mathrm{bus}}^{(m)}$ denote the regime-local admittance perturbation at node $m$. Using a minimal basis that spans the support of admissible network uncertainty, $\Delta\mathbf{Y}_{\mathrm{bus}}^{(m)} = \sum_{k=1}^{K_m} \delta y_k^{(m)} \,\mathbf{E}_k^{(m)}$. Uncertainty in $\Delta\mathbf{Y}_{\mathrm{bus}}^{(m)}$ is then characterized by the distribution of $\boldsymbol{\delta}_Y^{(m)}$, with conditional mean and covariance
\begin{equation}
    \mathbb{E}[\boldsymbol{\delta}_Y^{(m)} \mid \mathcal{R}_m] = \boldsymbol{\mu}_Y^{(m)}, \qquad
    \mathrm{Cov}(\boldsymbol{\delta}_Y^{(m)} \mid \mathcal{R}_m) = \boldsymbol{\Sigma}_Y^{(m)}.
\end{equation}

For a candidate branch $(i,j)$, let $Y_{ij}^{\mathrm{pred},(m)} = Y_{ij} + \Delta Y_{ij}^{(m)}$, with $\Delta Y_{ij}^{(m)} = \Delta G_{ij}^{(m)} + \mathrm{j}\Delta B_{ij}^{(m)}$. A near-zero admittance (disconnection) event is defined as

\begin{equation}
    E_{ij}^{(m)} = \left\{ \Big| Y_{ij}^{\mathrm{pred},(m)}\Big|^2 \le Y_{\min}^2 \right\},
\end{equation}
with $ \big| Y_{ij}^{\mathrm{pred},(m)}\big|^2 = \big(G_{ij} \!+\! \Delta G_{ij}^{(m)}\big)^2 \!+\! \big(B_{ij} \!+\! \Delta B_{ij}^{(m)}\big)^2$. Each connectivity event corresponds to a quadratic inequality in the jointly random variables $(\Delta G_{ij}^{(m)},\Delta B_{ij}^{(m)})$. Because these variables may be correlated with other perturbation elements, the resulting probability does not admit a closed-form expression and is evaluated via numerical integration.

The disconnection probability at node $m$ is $p_{ij}^{(m)} = \Pr\!\left(E_{ij}^{(m)} \mid \mathcal{R}_m\right)$, and the candidate set of connectivity regime events is
\begin{equation}
    \mathcal{C}_Y^{(m)} = \Big\{ (i,j) \,\,\Big|\,\, p_{ij}^{(m)} > \varepsilon_Y \Big\},
\end{equation}
for some significance threshold $\varepsilon_Y$. If $\mathcal{C}_Y^{(m)}=\varnothing$, no connectivity branching occurs at node $m$. Otherwise, the branch $(\widehat{i,j})\in\mathcal{C}_Y^{(m)}$ with the largest disconnection probability $p_{\hat{ij}}^{(m)}$ is selected for branching. Two candidate outcomes are defined:
\begin{align}
    E_{\mathrm{disc}}^{(m)} &= \left\{ \big|Y_{ij}^{\mathrm{pred},(m)}\big|^2 \le Y_{\min}^2 \right\} \mathrm{, and} \\
    E_{\mathrm{conn}}^{(m)} &= \left\{ \big|Y_{ij}^{\mathrm{pred},(m)}\big|^2 > Y_{\min}^2 \right\},
\end{align}
with regime-conditional edge probabilities
\begin{equation}
    p_{\mathrm{disc}\mid m} = p_{\hat{ij}}^{(m)}, \quad p_{\mathrm{conn}\mid m} = 1 - p_{\hat{ij}}^{(m)}.
\end{equation}
The corresponding node probabilities are $\gamma_{\mathrm{disc}}=\gamma_m \, p_{\mathrm{disc}\mid m}$ and $\gamma_{\mathrm{conn}}=\gamma_m \, p_{\mathrm{conn}\mid m}$.

If both resulting edge probabilities exceed the pruning threshold $\varepsilon_p$, two child nodes are created. In the disconnection child, the branch admittance is clamped to zero, $Y_{\hat{ij}}^{(\mathrm{disc})}=0$. The corresponding basis directions for $\Delta G_{\hat{ij}}$ and $\Delta B_{\hat{ij}}$ are removed from $\Delta\mathbf{Y}_{\mathrm{bus}}^{(m)}$, and the coefficient vector $\boldsymbol{\delta}_Y^{(m)}$ is reduced. The mean and covariance of the remaining coefficients are updated by restriction. If islanding occurs, the active system is reduced to the reachable sub-network. In the connected child, the network topology is preserved, and $\boldsymbol{\delta}_Y^{(m)}$ is conditioned on $E_{\mathrm{conn}}^{(m)}$. This yields a truncated distribution for $(\Delta G_{\hat{ij}}^{(m)},\Delta B_{\hat{ij}}^{(m)})$ while retaining all basis directions.

If only one outcome yields a node probability exceeding $\varepsilon_\gamma$, the parent regime is updated in place according to the surviving outcome. The complementary probability mass is collapsed to the corresponding connectivity event. If disconnection survives, the basis reduction and network update are applied as needed. If connection survives, the admittance uncertainty is truncated accordingly.

Let $\mathcal{B}_m$ denote the index set of basis directions active at node $m$, and let $\mathcal{C}_m \subseteq \mathcal{B}_m$ denote those corresponding to branch admittances that have already been tested for disconnection. A connectivity basis direction is considered resolved once its associated branch has been selected and evaluated for branching. A node $m$ is considered terminal when $\forall (i,j)\in\mathcal{B}_m\setminus\mathcal{C}_m,\quad p_{ij}^{(m)} \le \varepsilon_Y$, or when no further uncertain admittance directions remain.

\vspace{0.25cm}
\paragraph{$\Delta P$-Tree Construction}

At node $m$, let $\Delta\mathbf{P}_{\mathrm{inj}}^{(m)}\in\mathbb{R}^{|\mathcal{N}|}$ denote the active power injection components of the regime-local disturbance $\Delta\boldsymbol{\rho}^{(m)}$. Define the aggregation vector $\mathbf{h}=\mathbf{1}_{|\mathcal{N}|}$, so that the residual active power imbalance is
\begin{equation}
    \Delta P_{\mathrm{res}}^{(m)} = \mathbf{h}^\top \Delta\mathbf{P}_{\mathrm{inj}}^{(m)}
    = \sum_{i\in\mathcal{N}} \Delta P_i^{(m)},
\end{equation}
with $\Delta P_{\mathrm{res}}^{(0)}=\Delta P_{\mathrm{inj}}$ as in Section~\ref{sec:problem:response:active}. If $\Delta\mathbf{P}_{\mathrm{inj}}^{(m)}$ has conditional mean $\boldsymbol{\mu}_P^{(m)}$ and covariance $\boldsymbol{\Sigma}_P^{(m)}$, then
\begin{align}
    \mu_{\mathrm{res}}^{(m)} &= \mathbf{h}^\top \boldsymbol{\mu}_P^{(m)}, \\
    (\sigma_{\mathrm{res}}^{(m)})^2 &= \mathbf{h}^\top \boldsymbol{\Sigma}_P^{(m)}\,\mathbf{h}.
\end{align}

All participating generators respond to the same scalar imbalance. For generator $g$, $\Delta P_g^{(m)} = -\alpha_g^{(m)} \Delta P_{\mathrm{res}}^{(m)}$ with $\alpha_g^{(m)}>0$, and the signed margin to an active power limit is $M_{g,\mathrm{lim}}^{(m)} = s\big(P^{\mathrm{lim}} - (P_g + \Delta P_g^{(m)})\big)$. Substituting the participation response yields the scalar threshold condition
\begin{equation}
    M_{g,\mathrm{lim}}^{(m)} \le 0 \Longleftrightarrow
        s\,\Delta P_{\mathrm{res}}^{(m)} \le s\,\tau_{g,\mathrm{lim}}^{(m)},
\end{equation}
with $\tau_{g,\mathrm{lim}}^{(m)} = (P_g - P^{\mathrm{lim}})/\alpha_g^{(m)}$. Thus, all active power limit events correspond to parallel half-spaces in the scalar $\Delta P_{\mathrm{res}}^{(m)}$.

For each participating generator and limit, the conditional violation probability is
\begin{equation}
    p_{g,\mathrm{lim}}^{(m)} =
    \Pr\!\left(M_{g,\mathrm{lim}}^{(m)} \le 0 \,\middle|\, \mathcal{R}_m\right),
\end{equation}
and the candidate set of active power regime events is
\begin{equation}
    \mathcal{C}_P^{(m)} =
    \big\{ (g,\mathrm{lim}) : p_{g,\mathrm{lim}}^{(m)} > \varepsilon_P \big\},
\end{equation}
for threshold $\varepsilon_P$. If $\mathcal{C}_P^{(m)}=\varnothing$, no active power branching occurs at node $m$. Otherwise, the event $(\hat g,\widehat{\mathrm{lim}})$ with the largest $p_{g,\mathrm{lim}}^{(m)}$ is selected for branching.

Branching is induced by conditioning on the complementary events
\begin{equation}
    E_{\mathrm{lim}}^{(m)} =
    \big\{ s\,\Delta P_{\mathrm{res}}^{(m)} \le s\,\tau_{\hat g,\widehat{\mathrm{lim}}}^{(m)} \big\},
    \quad
    E_{\mathrm{int}}^{(m)} =
    \big(E_{\mathrm{lim}}^{(m)}\big)^c,
\end{equation}
with conditional probabilities 
\begin{equation}
    p_{\mathrm{lim}\mid m}=p_{\hat g,\widehat{\mathrm{lim}}}^{(m)}, \quad p_{\mathrm{int}\mid m}=1-p_{\hat g,\widehat{\mathrm{lim}}}^{(m)}.
\end{equation}

If both resulting edge probabilities exceed the pruning threshold $\varepsilon_p$, two child nodes are created. In the interior child, $\Delta P_{\mathrm{res}}^{(m)}$ is truncated to $E_{\mathrm{int}}^{(m)}$. In the limit child, generator $\hat g$ is clamped at $P_{\hat g}^{\widehat{\mathrm{lim}}}$, removed from the participation set, and participation factors are renormalized. The post-clamping imbalance is
\begin{equation}
    \Delta P_{\mathrm{res}}^{(\mathrm{lim})}
    \sim
    \big(\Delta P_{\mathrm{res}}^{(m)} \mid E_{\mathrm{lim}}^{(m)}\big)
    + \alpha_{\hat g}^{(m)} \tau_{\hat g,\widehat{\mathrm{lim}}}^{(m)},
\end{equation}
with subsequent branching using this updated distribution.

If only one outcome exceeds $\varepsilon_p$, the parent node is updated in place to represent the surviving regime, with the complementary probability mass collapsed into the corresponding saturation event. Here, $\Delta P_{\mathrm{res}}^{(m)}$ remains unchanged; it is only conditioned upon branching.

\vspace{0.25cm}
\paragraph{$\Delta Q$-Tree Construction}

The construction of the $\Delta Q$-tree follows the same sequential-pruning logic as the $\Delta P$-tree, differing in the structure of the reactive-power response and the physical consequences of saturation.

At node $m$, the regime-local disturbance $\Delta\boldsymbol{\rho}^{(m)}$ induces a linearized reactive-power response $\Delta \mathbf{Q}^{(m)} = \mathbf{S}_Q^{(m)}\,\Delta\boldsymbol{\rho}^{(m)}$, where $\mathbf{S}_Q^{(m)}$ depends on the current network topology, bus-type assignments, and set of voltage-controlling generators. For each generator $q$ operating in $PV$ mode, the predicted reactive-power output is $Q_q^{\mathrm{pred},(m)} = Q_q + \Delta Q_q^{(m)}$, with signed margin $M_{q,\mathrm{lim}}^{(m)} = s\big(Q^{\mathrm{lim}} - Q_q^{\mathrm{pred},(m)}\big)$.

Violation probabilities $p_{q,\mathrm{lim}}^{(m)}$, candidate sets $\mathcal{C}_Q^{(m)}$, event selection, and branching events $E_{\mathrm{lim}}^{(m)} = \{ M_{q,\mathrm{lim}}^{(m)} \le 0 \}$ and $E_{\mathrm{int}}^{(m)} = (E_{\mathrm{lim}}^{(m)})^c$ are defined the same way as in the $\Delta P$-tree.

In the limit child, generator $\hat{q}$ is clamped at $Q_{\hat{q}}^{\widehat{\mathrm{lim}}}$, and the bus is reclassified from $PV$ to $PQ$ if no other $PV$ generators remain at that bus. The linearized network equations are reconstructed, and a new sensitivity mapping $\mathbf{S}_Q^{(\mathrm{lim})}$ is formed before further branching. In the interior child, the generator remains in $PV$ mode and the disturbance distribution is conditioned on $E_{\mathrm{int}}^{(m)}$.

If only one outcome exceeds $\varepsilon_p$, the parent node is updated in place to represent the surviving regime. If the limit outcome survives, the $PV\!\rightarrow PQ$ transition is enforced and the complementary probability mass is collapsed into the violated limit. If the interior outcome survives, the disturbance distribution is truncated accordingly.

Branching continues until no remaining $PV$ generators exhibit $Q$-limit violation probabilities exceeding threshold $\varepsilon_Q$.

\vspace{0.25cm}
\paragraph{Mixture Representation and Evaluation of Leaf Moments and Probabilities}
The sequentially-pruned tree produces a finite set of leaf regimes $\{\mathcal{R}_\ell\}_{\ell=1}^{L}$ with associated unconditional weights $\gamma_\ell$ satisfying $\sum_{\ell=1}^{L}\gamma_\ell = 1$. Each leaf $\ell$ corresponds to a fully specified operating regime and an explicit conditioning region $\mathcal{R}_\ell$ obtained as the intersection of all edge events along the unique root-to-leaf path.

In leaf $\ell$, the regime-local disturbance is defined by the conditional law $\Delta\boldsymbol{\rho}\mid\mathcal{R}_\ell$, and deterministic regime edits which act on additive perturbation coordinates through projection, redistribution, or clamping operations. Since these operations are linear or constant-shift transformations on first-order deviation variables, their effect on the surviving disturbance can be represented by an affine transformation
\begin{equation}
    \Delta\boldsymbol{\rho}^{(\ell)}
    =
    \mathbf{A}_\ell\,\Delta\boldsymbol{\rho}
    +
    \mathbf{b}_\ell,
    \qquad
    \Delta\boldsymbol{\rho}\sim(\Delta\boldsymbol{\rho}\mid\mathcal{R}_\ell),
    \label{eq:rho_leaf_affine}
\end{equation}
where $\mathbf{A}_\ell$ encodes projection, redistribution, or reparameterization, and $\mathbf{b}_\ell$ accounts for fixed offsets introduced by clamping. Conditioning itself and the collapse of low-probability complementary outcomes into probability atoms act at the level of probability measures and are handled separately such that affine maps are applied
only to the propagated random component. The leaf-local moments are then
\begin{align}
    \boldsymbol{\mu}_\rho^{(\ell)}
        &= \mathbb{E}[\Delta\boldsymbol{\rho}^{(\ell)}]
         = \mathbf{A}_\ell\,\boldsymbol{\mu}_{\rho\mid\ell} + \mathbf{b}_\ell,
    \label{eq:rho_leaf_mean}\\
    \boldsymbol{\Sigma}_\rho^{(\ell)}
        &= \mathrm{Cov}(\Delta\boldsymbol{\rho}^{(\ell)})
         = \mathbf{A}_\ell\,\boldsymbol{\Sigma}_{\rho\mid\ell}\,\mathbf{A}_\ell^\top,
    \label{eq:rho_leaf_cov}
\end{align}
where $\boldsymbol{\mu}_{\rho\mid\ell} = \mathbb{E}[\Delta\boldsymbol{\rho}\mid\mathcal{R}_\ell]$ and $\boldsymbol{\Sigma}_{\rho\mid\ell} = \mathrm{Cov}(\Delta\boldsymbol{\rho}\mid\mathcal{R}_\ell)$. Probability mass accounting is handled entirely by the conditioning region $\mathcal{R}_\ell$, while the deterministic edits $(\mathbf{A}_\ell,\mathbf{b}_\ell)$ only modify the effective disturbance representation within the leaf.

For any monitored scalar quantity $x$, leaf $\ell$ contains a deterministic linear response map $\Delta x^{(\ell)} = \mathbf{S}_x^{(\ell)} \,\Delta\boldsymbol{\rho}^{(\ell)}$, with $\mathbf{S}_x^{(\ell)}$ computed from the physical model in that regime. Consequently, the induced leaf-local moments are
\begin{align}
    \mu_x^{(\ell)} &= \mathbb{E}[\Delta x^{(\ell)}]
        = \mathbf{S}_x^{(\ell)}\,\boldsymbol{\mu}_\rho^{(\ell)},
    \label{eq:leaf_mean_x}\\
    \sigma_{x}^{2,(\ell)} &= \mathrm{Var}(\Delta x^{(\ell)})
        = \mathbf{S}_x^{(\ell)}\,\boldsymbol{\Sigma}_\rho^{(\ell)}\,\mathbf{S}_x^{(\ell)\top}.
    \label{eq:leaf_var_x}
\end{align}
Define the leaf-wise violation probability as
\begin{equation}
    p_{\mathrm{viol},x}^{(\ell)}
    := \Pr\!\left(s\,\Delta x^{(\ell)} \ge M_x^\star\right),
    \label{eq:leaf_viol_prob}
\end{equation}
with $p_{\mathrm{rel},x}^{(\ell)} = 1 - p_{\mathrm{viol},x}^{(\ell)}$. The overall probabilities are obtained
by mixture aggregation over leaves,
\begin{equation}
    p_{\mathrm{viol},x} = \sum_{\ell=1}^{L} \gamma_\ell\,p_{\mathrm{viol},x}^{(\ell)}, \quad
    p_{\mathrm{rel},x} = \sum_{\ell=1}^{L} \gamma_\ell\,p_{\mathrm{rel},x}^{(\ell)}.
    \label{eq:mixed_probs}
\end{equation}

Lastly, in each node regime, we can compute $\bar{\mathbf{z}}^{(l)}$ for evaluating quasi-duals and define mixture-model quasi-duals over the tree over the union or intersection of active constraint sets, as desired. We omit these details for brevity.

\vspace{0.25cm}
\paragraph{Combinatorial Reduction}
Sequential pruning constructs a reduced binary regime tree whose growth is tied to the number of regime events realized along each path. In the worst case, the number of leaf regimes is bounded by
\begin{equation}
    R \;\le\; 2^{\,N_P + N_{\Delta Y} + N_Q}.
\end{equation}
In practice, the realized tree is far smaller. Only events with conditional violation probabilities exceeding prescribed tolerances are admitted as branching candidates, suppressing growth driven by distribution tails. Conditioning at each interior node further reduces the likelihood of downstream events, while clamped outcomes rapidly diminish the remaining probability mass. Moreover, each $PV\!\to PQ$ transition permanently removes a generator from subsequent reactive-power consideration.

If additional reduction is required, further approximations may be applied. For example, pruning can be based on cumulative node probability $\gamma_n$ rather than conditional probabilities $p_{n|m}$, or enforced through structural limits on tree breadth, depth, or total node count. Such choices trade enforcement of hard constraints on generators for computation time. We recommend always checking for islanding.

\vspace{0.25cm}
\subsubsection{Approximation of Distribution-Free Cantelli Bounds in an $\alpha$-Stressed Regime}
\label{sec:probabilistic:alpha_regime}

The sequentially-pruned regime-tree above is feasible when the distribution of $\Delta\boldsymbol{\rho}$ is specified. When only first and second moments of $\Delta\boldsymbol{\rho}$ are available, neither the node probabilities $\gamma_\ell$ nor the conditional probabilities $p_{n|m}$ can be identified uniquely, since many distinct distributions can share the same mean and covariance but induce different probabilities of regime transitions.

To obtain a tractable, distribution-free alternative, we construct a single \emph{$\alpha$-stressed regime}. Using only first and second moments we identify which regime changes are unavoidable at a prescribed confidence level $\alpha$, and deterministically enforce those changes prior to evaluating constraint margins. All remaining constraints are then assessed within this single $\alpha$-stressed regime using Cantelli bounds. This construction yields conservative guarantees while avoiding combinatorial regime enumeration. This method can be applied even if the distribution of $\Delta\boldsymbol{\rho}$ is known.

Starting from the base operating point and disturbance moments $(\boldsymbol{\mu}_\rho,\boldsymbol{\Sigma}_\rho)$, the $\alpha$-stressed regime approximation proceeds in three stages, following the same ordering of regime classes as the deterministic and mixture-based constructions. Each stage applies an $\alpha$-confidence test to all candidate events of that class simultaneously. All tests that cannot be certified at confidence $\alpha$ are conservatively enforced. The system model, disturbance representation, and linearized response maps are then updated deterministically before proceeding to the next stage.

\vspace{0.25cm}
\paragraph{$\Delta \mathbf{Y}_{\mathrm{bus}}$ $\alpha$-Test}
For each branch $(i,j)$ with uncertain admittance, let
\begin{equation}
    \mathbf{Y}_{ij}^{\mathrm{pred}} =
    \begin{bmatrix}
        G_{ij} + \Delta G_{ij} \\
        B_{ij} + \Delta B_{ij}
    \end{bmatrix}
\end{equation}
be the predicted admittance vector. Its mean and covariance, $(\boldsymbol{\mu}_{ij},\boldsymbol{\Sigma}_{ij})$, are obtained as affine projections of $(\boldsymbol{\mu}_\rho,\boldsymbol{\Sigma}_\rho)$. Disconnection is defined by the event $\|\mathbf{Y}_{ij}^{\mathrm{pred}}\|_2 \le Y_{\min}$.

With only first and second moments, this quadratic event cannot be evaluated exactly. Instead, we create a distribution-free $\alpha$-confidence certificate. Using the expectation of a quadratic form \cite{1981_trace_quadratic}, we have
\begin{equation}
    \mathbb{E}\!\left[\left\|\mathbf{Y}_{ij}^{\mathrm{pred}} - \boldsymbol{\mu}_{ij}\right\|_2^2\right] =
         \mathrm{tr}(\boldsymbol{\Sigma}_{ij}).
\end{equation}
Applying Markov's inequality \cite{2019_markov_nonnegative} to $\|\mathbf{Y}_{ij}^{\mathrm{pred}}-\boldsymbol{\mu}_{ij}\|_2^2$ gives
\begin{equation}
    \Pr\!\left( \left\|\mathbf{Y}_{ij}^{\mathrm{pred}} - \boldsymbol{\mu}_{ij}\right\|_2^2
        \ge \frac{\mathrm{tr}(\boldsymbol{\Sigma}_{ij})}{1-\alpha} \right) \le 1-\alpha,
\end{equation}
or equivalently,
\begin{equation}
    \Pr\!\left( \left\|\mathbf{Y}_{ij}^{\mathrm{pred}} - \boldsymbol{\mu}_{ij}\right\|_2 \le r_{\alpha,ij} \right) \ge \alpha, \quad 
    r_{\alpha,ij} = \sqrt{\frac{\mathrm{tr}(\boldsymbol{\Sigma}_{ij})}{1-\alpha}}.
    \label{eq:yij_ball_confidence}
\end{equation}
Thus, with confidence at least $\alpha$, the random vector $\mathbf{Y}_{ij}^{\mathrm{pred}}$ lies inside the
Euclidean ball of radius $r_{\alpha,ij}$ around its mean. By the triangle inequality \cite{2001_triang_inequal},
\begin{align}
    \|\boldsymbol{\mu}_{ij}\|_2 &= \left\|\mathbf{Y}_{ij}^{\mathrm{pred}} - (\mathbf{Y}_{ij}^{\mathrm{pred}}-\boldsymbol{\mu}_{ij})\right\|_2 \nonumber \\
    &\le \|\mathbf{Y}_{ij}^{\mathrm{pred}}\|_2 + \|\mathbf{Y}_{ij}^{\mathrm{pred}}-\boldsymbol{\mu}_{ij}\|_2, 
    \nonumber \\
    \implies \|\mathbf{Y}_{ij}^{\mathrm{pred}}\|_2 &\ge \|\boldsymbol{\mu}_{ij}\|_2 - \|\mathbf{Y}_{ij}^{\mathrm{pred}}-\boldsymbol{\mu}_{ij}\|_2. 
    \label{eq:triangle_rearranged}
\end{align}
Combining \eqref{eq:triangle_rearranged} with the $\alpha$-confidence event in \eqref{eq:yij_ball_confidence}, with probability at least $\alpha$,
\begin{equation}
    \|\mathbf{Y}_{ij}^{\mathrm{pred}}\|_2
    \ge
    \|\boldsymbol{\mu}_{ij}\|_2 - r_{\alpha,ij}.
    \label{eq:yij_norm_lower_confidence}
\end{equation}
We thus define the \emph{$\alpha$-confidence connectivity margin}
\begin{equation}
    M_{\alpha,Y_{ij}} = \|\boldsymbol{\mu}_{ij}\|_2 - 
        \sqrt{\tfrac{\mathrm{tr}(\boldsymbol{\Sigma}_{ij})}{1-\alpha}} - Y_{\min}.
\end{equation}
If $M_{\alpha,Y_{ij}}>0$, branch $(i,j)$ is certified to remain connected with confidence at least $\alpha$. Otherwise, disconnection cannot be excluded at level $\alpha$ and the branch is removed.

All branches with $M_{\alpha,Y_{ij}}\le 0$ are simultaneously disconnected. The network is then reduced to the subgraph reachable from the slack bus. Corresponding admittance uncertainty directions are removed from $\Delta\boldsymbol{\rho}$, and the disturbance moments $(\boldsymbol{\mu}_\rho,\boldsymbol{\Sigma}_\rho)$ are updated by restriction to the remaining degrees of freedom. The network model and linearization are reconstructed before proceeding.

\vspace{0.25cm}
\paragraph{$\Delta P$ $\alpha$-Test}
Let $\Delta P_{\mathrm{res}}=\mathbf{h}^\top\Delta\mathbf{P}_{\mathrm{inj}}$ denote the net active power imbalance, with mean $\mu_{\mathrm{res}}$ and variance $\sigma_{\mathrm{res}}^2$. For each participating generator $g$ with participation factor $\alpha_g$, the predicted response is $\Delta P_g=-\alpha_g\Delta P_{\mathrm{res}}$. For each active power limit, define the signed margin $M_{g,\mathrm{lim}}^\star$ as in the deterministic case. Using the moment-based $\alpha$-confidence margin, $M_{\alpha,g,\mathrm{lim}} = M_{g,\mathrm{lim}}^\star - s\,\mu_{g} - \xi_\alpha\,\sigma_{g}$, all generators for which $ M_{\alpha,g,\mathrm{lim}}\le 0$ are clamped at the corresponding limit.

Let $\mathcal{G}_{\mathrm{sat}}^{(P)}$ denote the set of clamped generators. These are removed from the participation set, and participation factors are renormalized over the remaining generators. The residual imbalance is updated affinely by subtracting the deterministic contributions of the clamped units, which shifts $\mu_{\mathrm{res}}$ but leaves $\sigma_{\mathrm{res}}$ unchanged. The active power control model and linear response mappings are then updated.

\vspace{0.25cm}
\paragraph{$\Delta Q$ $\alpha$-Test}
For each generator operating in $PV$ mode, the linearized reactive power response is $\Delta Q_q = \mathbf{S}_{q}\,\Delta\boldsymbol{\rho}$, with moments $(\mu_{q},\sigma_{q})$ computed from the current disturbance moments. For each reactive power limit, define the $\alpha$-confidence margin $M_{\alpha,q,\mathrm{lim}} = M_{q,\mathrm{lim}}^\star - s\,\mu_{q} - \xi_\alpha\,\sigma_{q}$. All generators with $M_{\alpha,q,\mathrm{lim}} \le 0$ are clamped at their reactive limits and reclassified from $PV$ to $PQ$ mode. The bus-type assignment is updated, the linearized network equations are reconstructed, and the sensitivities are recomputed.

\vspace{0.25cm}
\paragraph{Evaluation in the $\alpha$-Stressed Regime.}
After applying all three classes of $\alpha$-tests, the system is left in a single $\alpha$-stressed operating regime with a fully specified network topology, generator participation set, and bus-type classification. The disturbance representation $\Delta\boldsymbol{\rho}$ has updated moments $(\boldsymbol{\mu}_\rho^{(\alpha)}, \boldsymbol{\Sigma}_\rho^{(\alpha)})$, and all linear response mappings $\mathbf{S}_x^{(\alpha)}$ are recomputed consistently with this regime.

For any $x$, $\Delta x = \mathbf{S}_x^{(\alpha)}\,\Delta\boldsymbol{\rho}$, with moments
\begin{equation}
    \mu_x^{(\alpha)} = \mathbf{S}_x^{(\alpha)}\boldsymbol{\mu}_\rho^{(\alpha)}, \quad
    (\sigma_x^{(\alpha)})^2 = \mathbf{S}_x^{(\alpha)} \boldsymbol{\Sigma}_\rho^{(\alpha)} \mathbf{S}_x^{(\alpha)\top}.
\end{equation}
Violation and release probabilities are then bounded using Cantelli’s inequality with these moments, yielding distribution-free guarantees that are valid for all disturbance distributions consistent with the specified first and second moments.

\begin{figure*}[!t]
    \centering
    \begin{subfigure}[t]{0.48\textwidth}
        \centering
        \includegraphics[width=\linewidth]{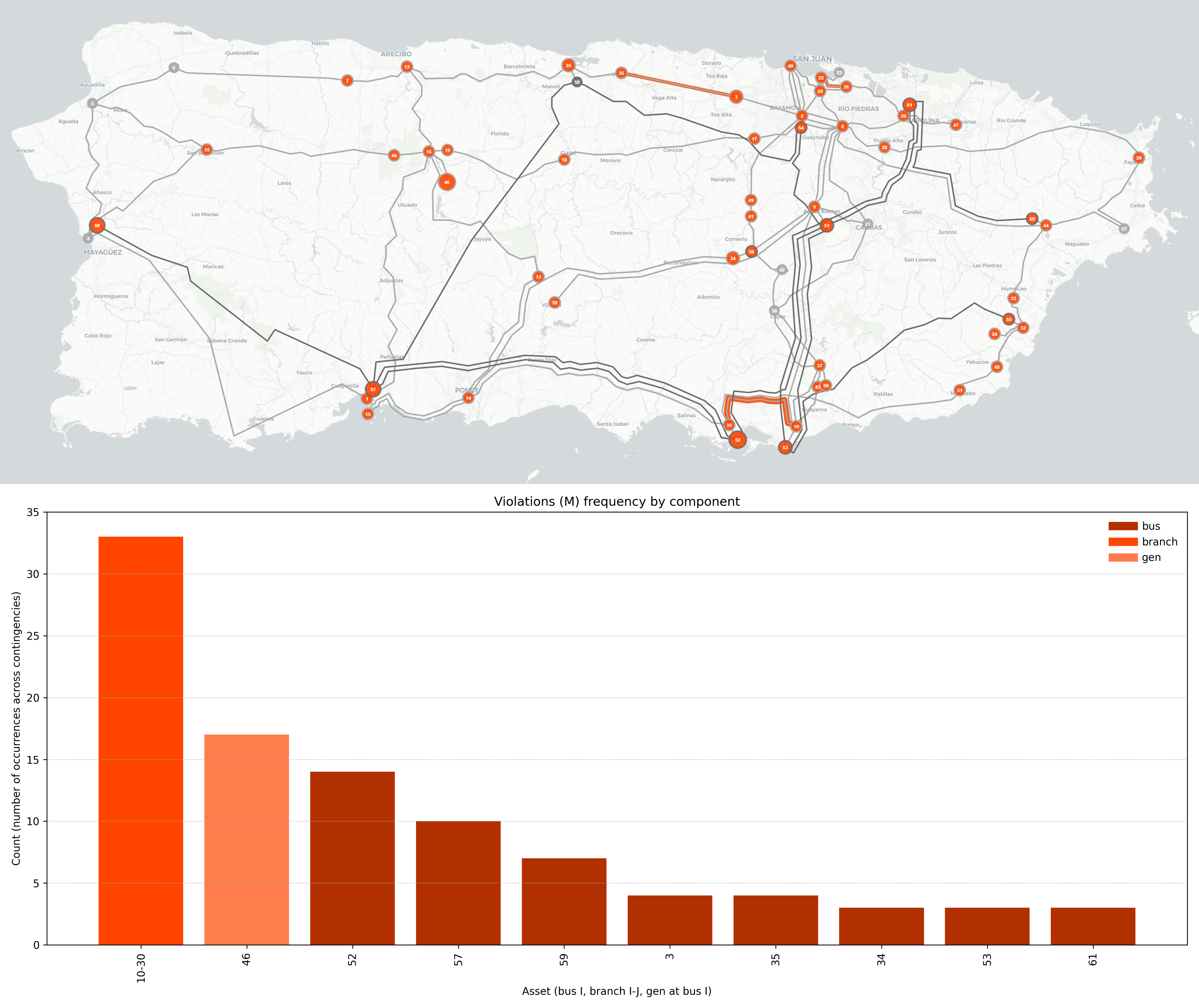}
        \caption{$N\!-\!1$ screening with preset voltage setpoints.}
        \label{fig:pr_preset_V}
    \end{subfigure}\hfill
    \begin{subfigure}[t]{0.48\textwidth}
        \centering
        \includegraphics[width=\linewidth]{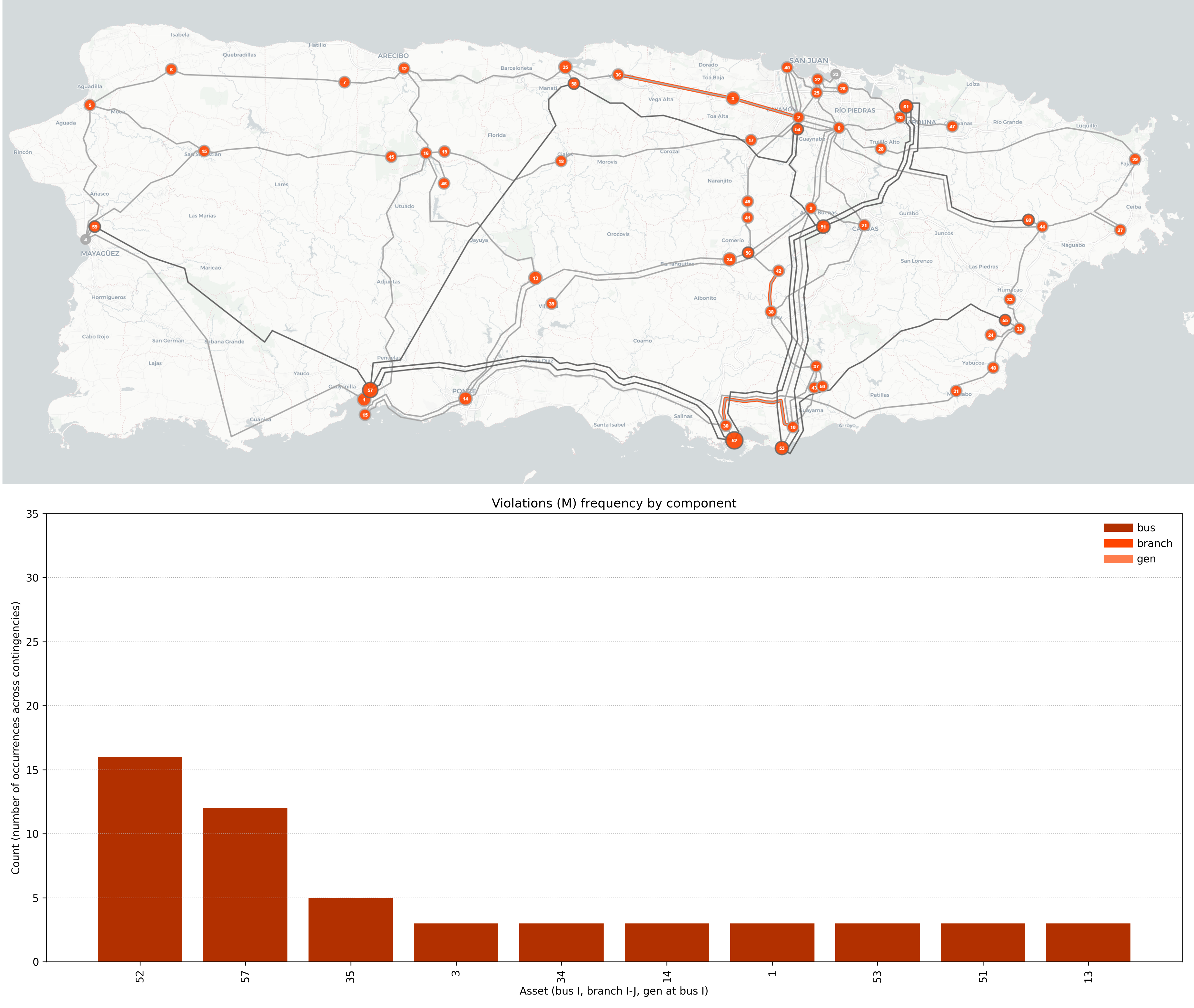}
        \caption{$N\!-\!1$ screening with optimized voltage setpoints.}
        \label{fig:pr_optimized_V}
    \end{subfigure}
    \vspace{4pt}
    \caption{$N\!-\!1$ contingency screening on Puerto Rican bulk power system model, around a base operating point representing peak load conditions with dispatch optimized using AC OPF economic dispatch with (a) preset and (b) optimized voltage setpoints.}
    \label{fig:pr_preset_vs_optimized}
    \vspace{-10pt}
\end{figure*}

\subsection{Robustness Screening and Meaningful Uncertainty Models}
\label{sec:disturb}

Sections~\ref{sec:probabilistic:linear}--\ref{sec:probabilistic:regimes} develop a framework that is distribution-agnostic. In practice, system-wide screening requires disturbance models that reflect credible operational uncertainty. Here we briefly review commonly used disturbance models.

\vspace{0.25cm}
\paragraph{Independent Uncertainties}
Independent disturbances across injections and branches take the block diagonal form
\begin{equation}
    \Delta\boldsymbol{\rho} \sim \mathcal{P}\!\left(\boldsymbol{\mu},\mathrm{diag}(\Sigma_1^2,\ldots,\Sigma_{N_G+N_Y}^2)\right)
\end{equation}
and have been explored extensively in both PLF and CC-OPF contexts \cite{2006_independent_injections, 2021_independent_loads, 2018_independent_gaussian_wind}. Such models define Cartesian uncertainty sets and naturally support $k^{\mathrm{th}}$-order screening analyses, analogous to $N\!-\!k$ screening. For known distributions, a system is $N+\mathcal{P}(\boldsymbol{\mu},\boldsymbol{\Sigma})_k$ robust if all admissible $k^{\mathrm{th}}$-order $\Delta\boldsymbol{\rho}$ satisfy prescribed violation and release probability tolerances. When only first and second moments are available, an analogous $N+(\boldsymbol{\mu},\boldsymbol{\Sigma})_k$ notion applies.

\vspace{0.25cm}
\paragraph{Load and Renewable Uncertainty}
Load, wind, and solar photovoltaic uncertainties exhibit strong temporal and spatial dependence driven by shared factors such as weather conditions, time of day, and geographic proximity. The literature captures these correlations using joint models, including copula-based methods, mixed PDF/CDF formulations, Gaussian mixture models, and sampling approaches \cite{2014_wind_modeling_review, 2020_pfl_wind_solar, 2025_review_uncertainty_models}. These joint models induce correlated disturbance directions in $\Delta\boldsymbol{\rho}$, which can be sampled or approximated within the proposed screening framework without modification.

\vspace{0.25cm}
\paragraph{Line Parameter and Topology Uncertainty}
Transmission uncertainty arises from errors in line and transformer parameters, operating conditions, and network switching or outages. In the PLF literature, line parameters are treated as uncertain inputs to the power-flow equations using probabilistic, bounded, or scenario-based models, while topology uncertainty is handled through discrete contingencies or outage probabilities \cite{2017_plf_review}. Here, continuous parameter uncertainty enters through perturbations of $\mathbf{Y}_{\mathrm{bus}}$ under fixed topology, whereas discrete topology changes are handled explicitly through the regime and mixture constructions of Section~\ref{sec:probabilistic:regimes}.

\begin{figure*}
    \centering
    \begin{subfigure}[t]{0.49\textwidth}
        \centering
        \includegraphics[width=\linewidth]{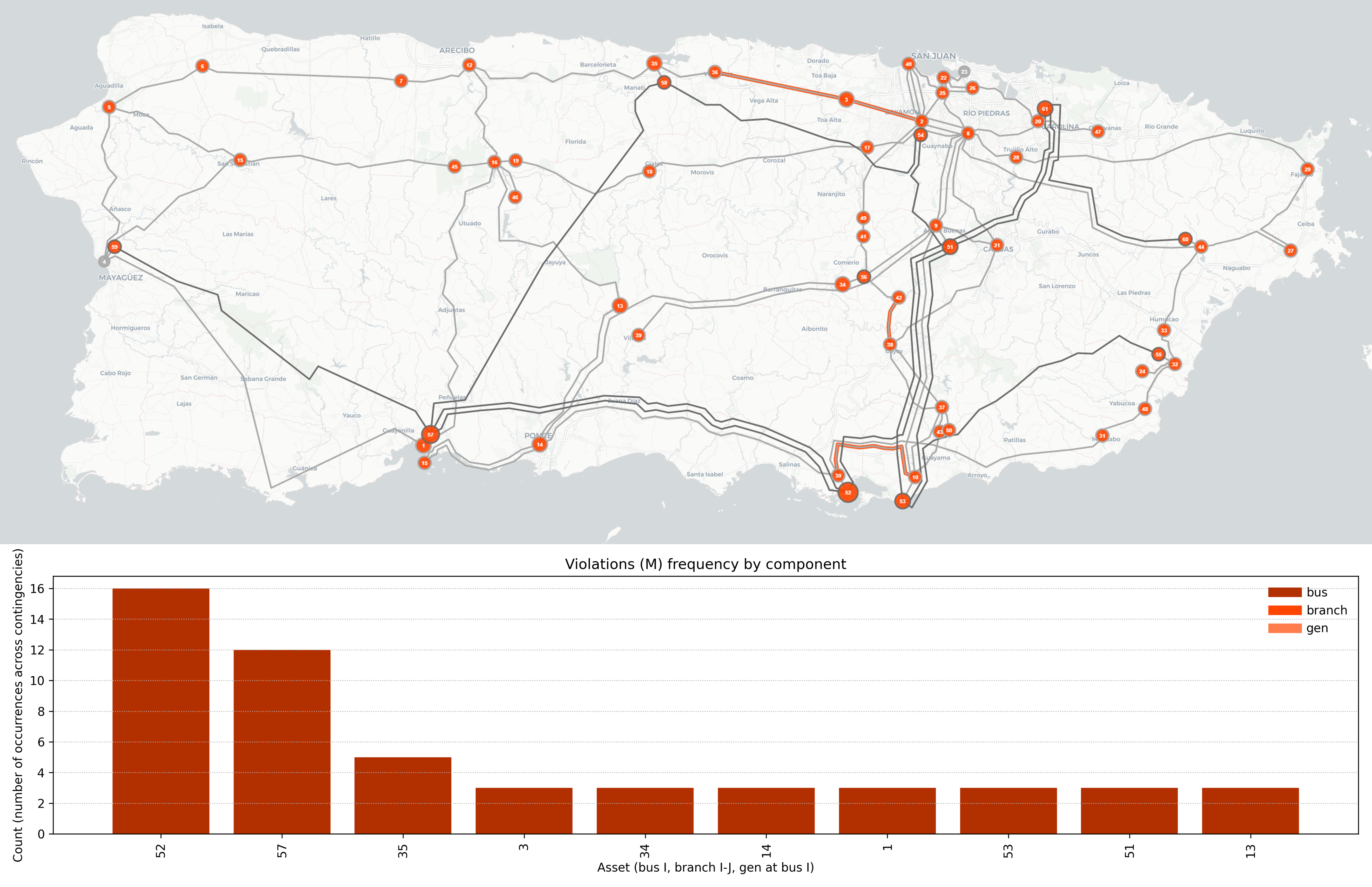}
        \caption{Limit violation ($M_x<0$) frequencies from $N\!-\!1$ screening.}
        \label{fig:Nm1}
    \end{subfigure}\hfill
    \begin{subfigure}[t]{0.49\textwidth}
        \centering
        \includegraphics[width=\linewidth]{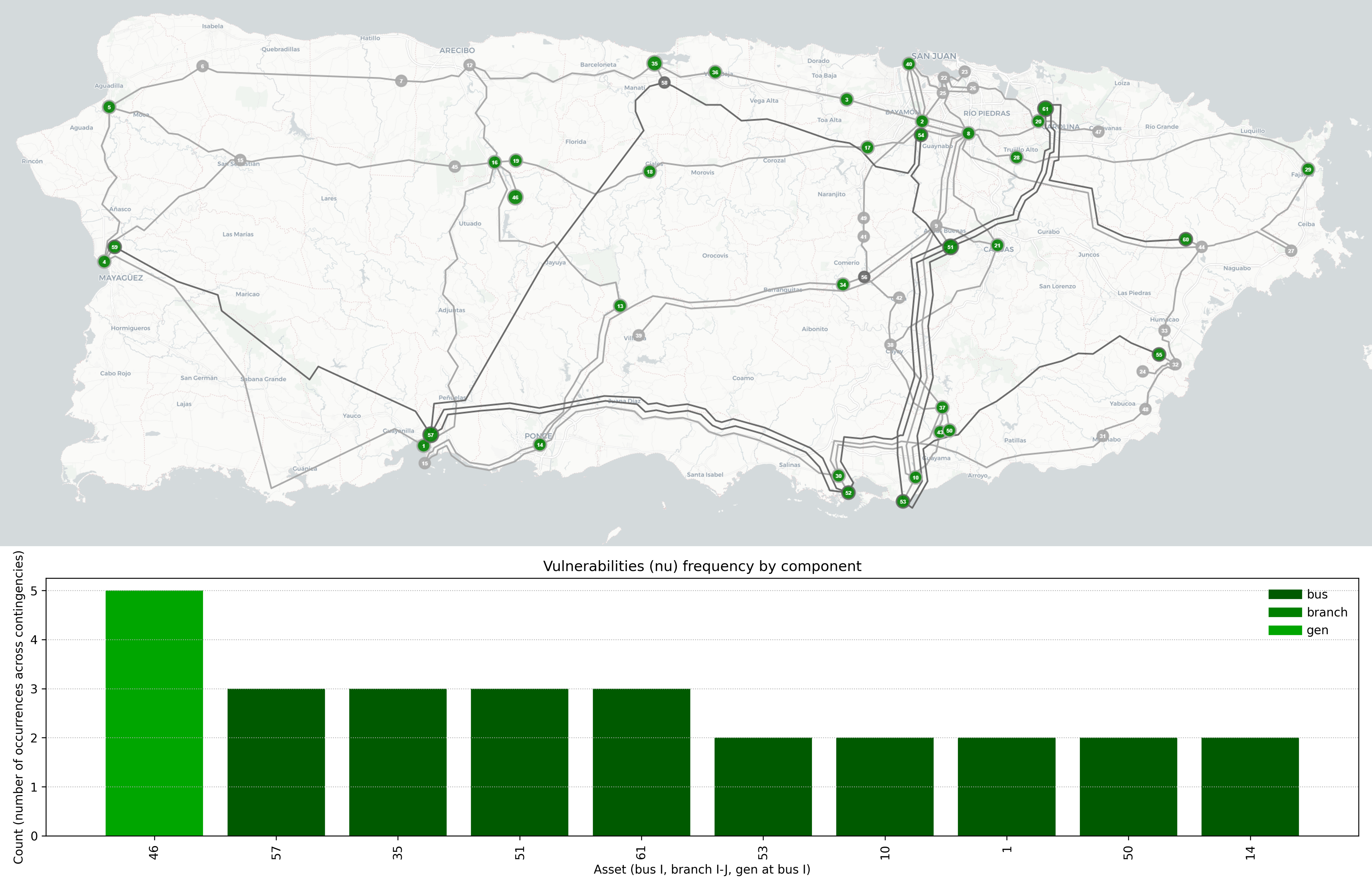}
        \caption{Quasi-dual vulnerabilities ($\tilde\nu_x>0$) from $C\!-\!1$ screening.}
        \label{fig:Cm1}
    \end{subfigure}

    \begin{subfigure}[t]{0.49\textwidth}
        \centering
        \includegraphics[width=\linewidth]{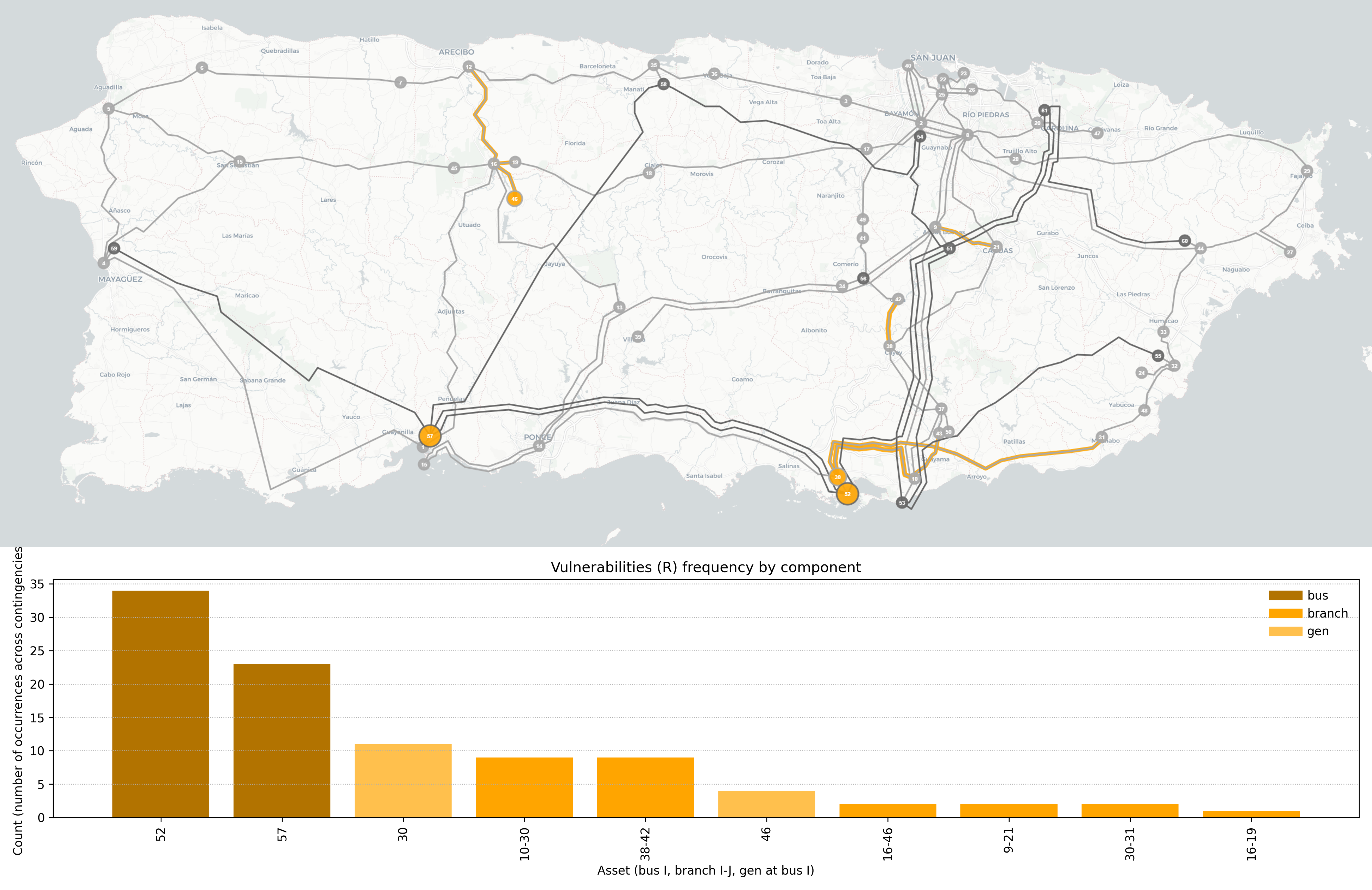}
        \caption{Marginal vulnerabilities ($R_x\in[-10,0)$) from $N\!+\!\delta_1$ screening.}
        \label{fig:Npd1}
    \end{subfigure}\hfill
    \begin{subfigure}[t]{0.49\textwidth}
        \centering
        \includegraphics[width=\linewidth]{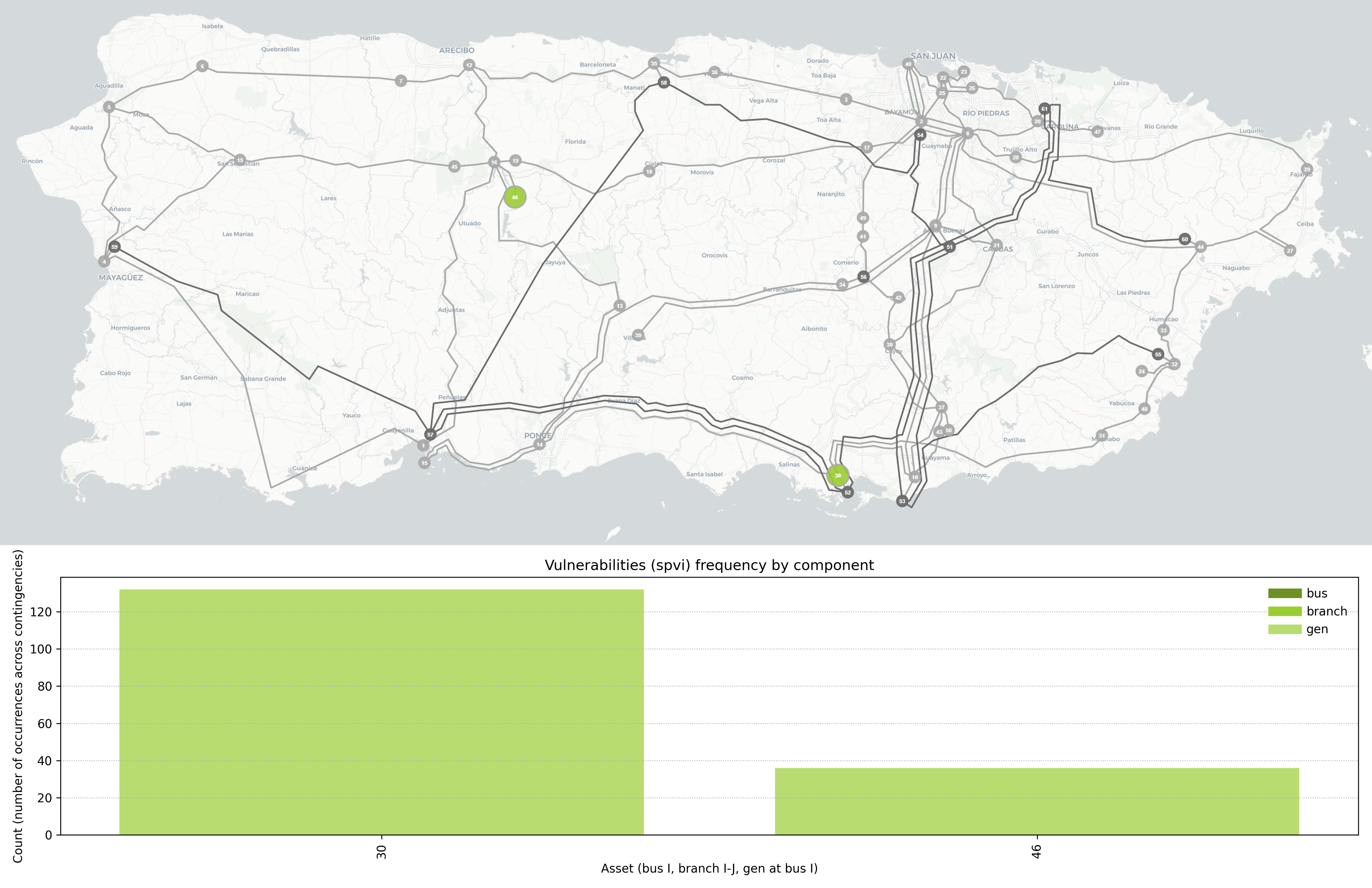}
        \caption{SPVI vulnerabilities ($\varsigma_x>0$) from $C\!+\!\delta_1$ screening.}
        \label{fig:Cpd1}
    \end{subfigure}

    \begin{subfigure}[t]{0.49\textwidth}
        \centering
        \includegraphics[width=\linewidth]{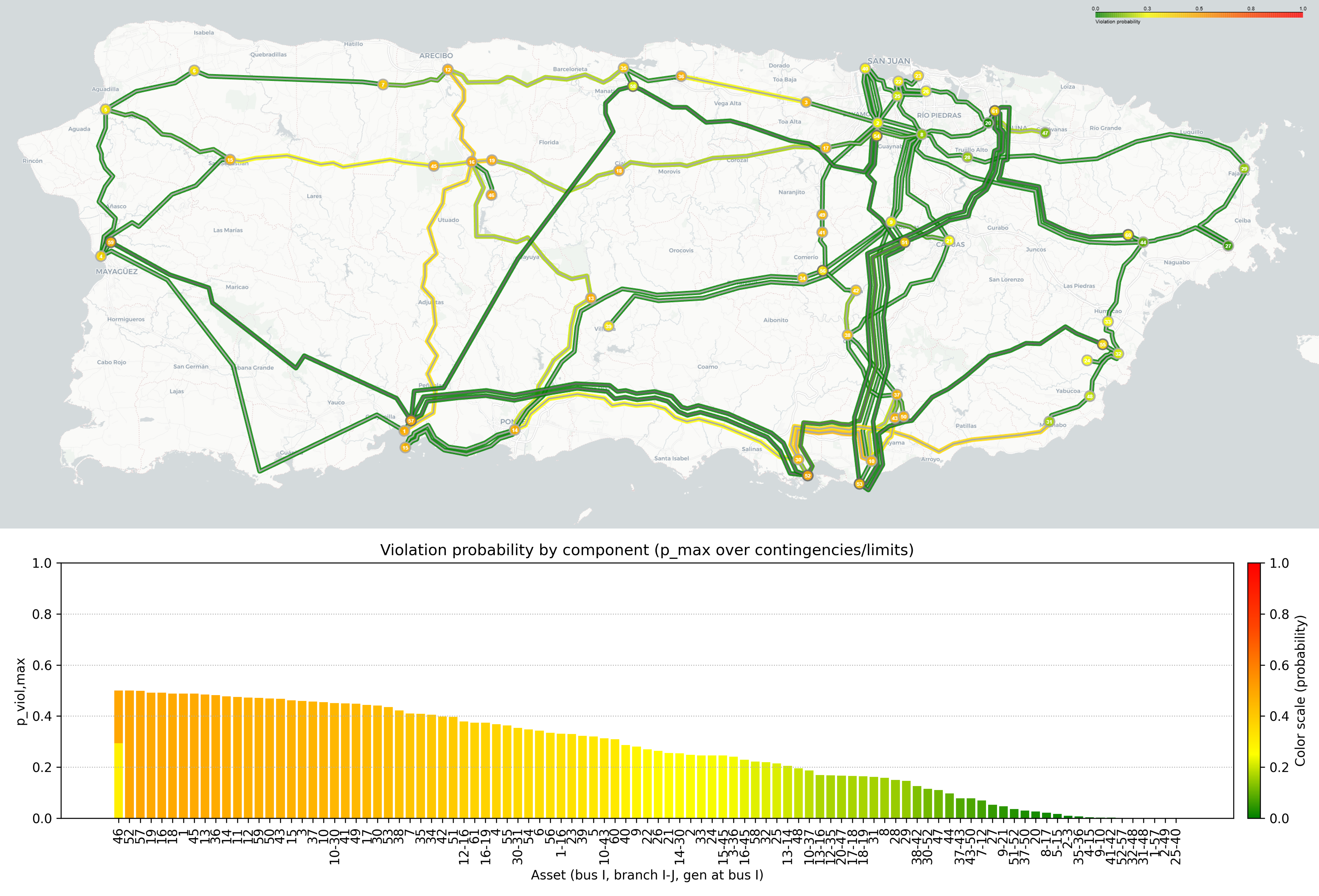}
        \caption{Max violation probabilities from $N\!+\!\mathcal{N}(\boldsymbol{\mu}, \boldsymbol{\Sigma})_1$ screening.}
        \label{fig:NpNmS1}
    \end{subfigure}\hfill
    \begin{subfigure}[t]{0.49\textwidth}
        \centering
        \includegraphics[width=\linewidth]{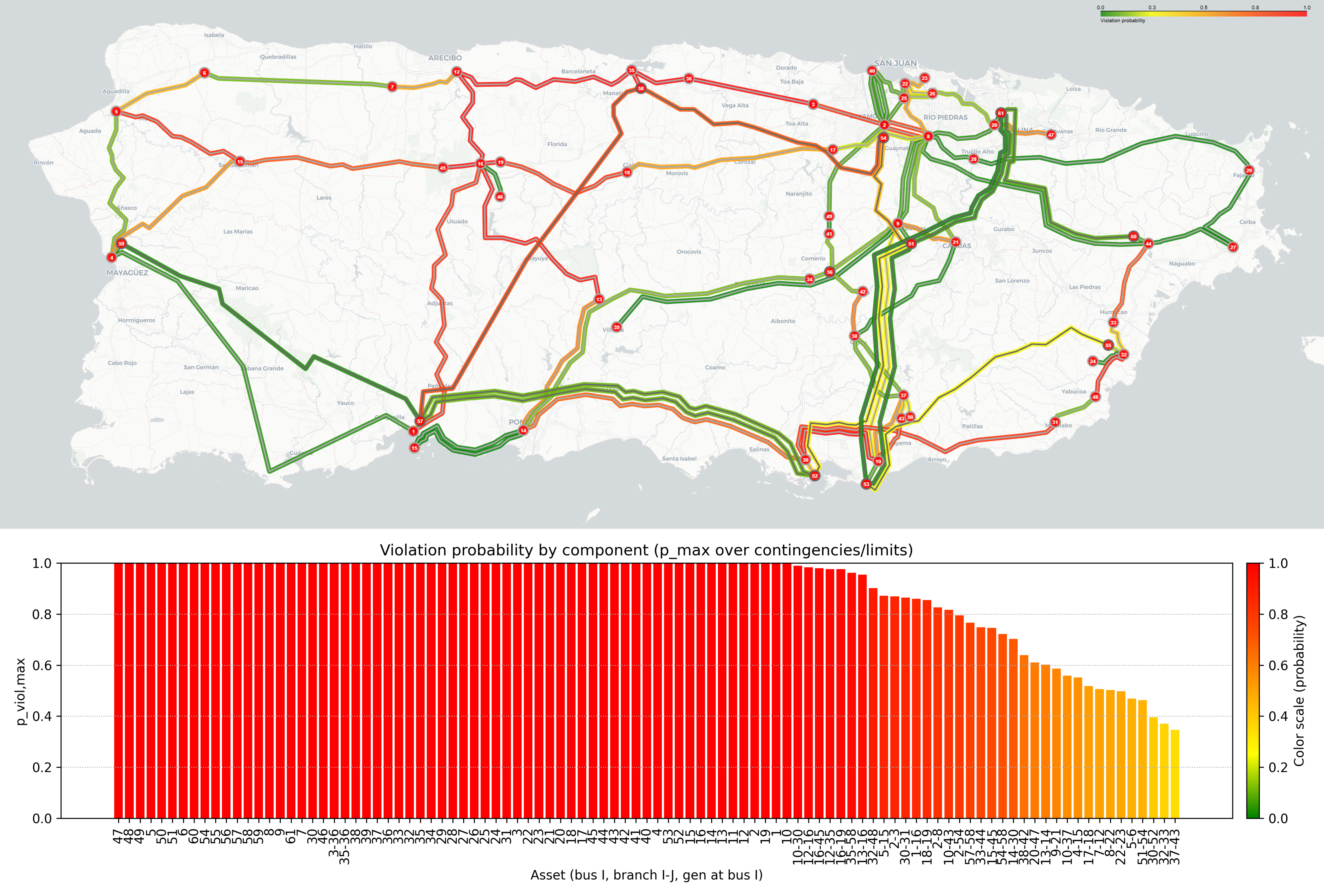}
        \caption{Max violation probabilities from $N\!+\!(\boldsymbol{\mu}, \boldsymbol{\Sigma})_1$ screening.}
        \label{fig:NpmS1}
    \end{subfigure}
    \caption{Geospatial visualizations of system-wide screening metrics on the Puerto Rican bulk power system under a peak-load operating point solved by AC OPF with optimized voltage setpoints. Panels (a)–(d) show deterministic $N\!-\!1$, $C\!-\!1$, $N\!+\!\delta(1)$, and $C\!+\!\delta(1)$ screenings, with the ten components exhibiting the highest frequencies highlighted. Panels (e)–(f) present probabilistic single-contingency screening under Gaussian disturbances using the sequentially pruned mixture model and the $\alpha$-stressed regime method, respectively. For probabilistic panels, the 100 largest violation probabilities are displayed, with generator probabilities aggregated to their corresponding buses.}
    \label{fig:pr_dashboard}
    \vspace{-10pt}
\end{figure*}

\section{Case Study: Demonstration on Puerto Rican Bulk Power System} 
\label{sec:casestudy}

To demonstrate how the proposed screening metrics can be communicated at scale, we apply them to a publicly available AC power flow model of Puerto Rico's bulk power system (BPS) first created by \cite{2019_pr_ilic} and visualize results geospatially. The model has 12 generators, 960 loads, 385 buses and 861 lines representing the 38~kV, 115~kV and 230~kV network, derived from public data sources. We implement minor modifications to limits, setpoints, and branch impedances as described in \cite{2024_sm_thesis}. The 38~kV sub-transmission level is approximated using Kron reduction. As such, all reported violations, contingencies, and screening visualizations are restricted to the 115~kV and 230~kV BPS layer. Puerto Rico's BPS spans the island with major load centers including San Juan, Ponce, Mayag\"uez, Arecibo, and Caguas, while legacy thermal generation is largely concentrated near coastal ports. Bulk transfers commonly traverse the mountainous interior \cite{2024_sm_thesis}.

We consider a peak loading scenario, which is solved in \cite{2024_sm_thesis} with an AC OPF economic dispatch formulation, using the \emph{SmartGridz} solver \cite{2025_smartgridz_manual} with both preset and adjustable voltage setpoints at $PV$ buses. All generators are participating in $P$-balancing, in proportion to their MW capacities.

First, we perform standard $N\!-\!1$ screening on both cases with the settings mentioned, considering generator outages and outages on 115~kV branches and above. Figure~\ref{fig:pr_preset_vs_optimized} shows violation frequencies on buses, branches, and generators in both cases, illustrated geospatially (generators are aggregated to corresponding buses) and as a barplot for the top 10 components with the most limit violations. The side-by-side comparison shows that by optimizing voltage setpoints, the number of violations system-wide decreases and stress is more evenly distributed across components.


We next extend the analysis to compare the array of screening tools presented in this paper, on the optimized voltage setpoint case. In particular, we perform:
\begin{enumerate}[itemsep=0pt, topsep=2pt]
    \item $N\!-\!1$ screening as above.
    \item $C\!-\!1$ screening via the $N\!-\!1$ perturbed states, marking all positive inequality quasi-duals as vulnerable.
    \item $N+\delta_1$ screening, using aligned disturbance directions on the $N\!-\!1$ component set, and $\varepsilon_R=10$.
    \item $C+\delta_1$ screening, using the $N+\delta_1$ directions, marking all positive inequality SPVI indices as vulnerable.
    \item $N\!+\!\mathcal{N}(\boldsymbol{\mu}, \boldsymbol{\Sigma})_1$ screening, using multivariate Gaussian distributions with zero means and 0.02\% of base injections and admittances as variances on the $N\!-\!1$ component set. We set $Y_{\min}=1e^{-5}$ and an equal significance threshold of 5\% for all condition tests. We draw 10,000 Monte Carlo samples for each instantiation of $\Delta\boldsymbol{\rho}^{(m)}$.
    \item $N\!+\!(\boldsymbol{\mu}, \boldsymbol{\Sigma})_1$ screening, using the same moments as in the $N\!+\!\mathcal{N}(\boldsymbol{\mu}, \boldsymbol{\Sigma})_1$ screening. We consider $\alpha=95\%$ confidence tests and Cantelli bounds.
\end{enumerate}

The results are shown in Figure~\ref{fig:pr_dashboard} as a dashboard. All maps presented in this paper are screenshots of interactive maps that identify which uncertainties correspond to which exceedance flags or probabilities per component. 

Each screening result highlights different aspects of system robustness. $N\!-\!1$ and $C\!-\!1$ screens (Figures~\ref{fig:Nm1}–\ref{fig:Cm1}) correspond to classical contingency analyses with regime transitions. They identify components prone to limit violations under outages and reveal how frequently active constraints exert positive economic pressure via quasi-duals. In contrast, $N\!+\!\delta_1$ and $C\!+\!\delta_1$ screens (Figures~\ref{fig:Npd1}–\ref{fig:Cpd1}) provide local analogs that highlight components which are directionally vulnerable to violations or shadow-price amplification.


The deterministic results are given using frequency of threshold exceedance, but could be shown instead using continuous measure such as mean or median magnitudes of exceedance. We implement this with the probabilistic results in Figure~\ref{fig:NpNmS1} and Figure~\ref{fig:NpmS1} by showing violation probabilities. Each plot shows, given uncertainty sets with the same first and second moments, the maximum probability of violation per component over the uncertainty sets. As expected, the results align, with the $\alpha$-stressed regime method systematically providing more conservative estimates.

Interestingly, across all screening methods, the system was found to be most vulnerable to bus voltage violations.

\section{Conclusion}
\label{sec:conclusion}

This paper presented a unified diagnostic framework for assessing the robustness of AC operating points, by perturbing primal and dual states derived from AC OPF solutions. We develop a control- and constraint-aware steady-state response model to assess how operating points change in response to injection and topology perturbations, while considering regime changes driven by connectivity loss, generator saturation, and $PV\!\rightarrow\!PQ$ transitions. Propagating disturbances through a common response mapping enables consistent deterministic and probabilistic robustness metrics that generalize classical $N\!-\!1$ screening to include economic indicators, local neighborhoods, and uncertain inputs.

To interpret economic stress away from optimality, we introduced quasi-dual variables as local geometric analogs of shadow prices, which are nonzero for active constraints, similar to dual variables derived from optimization. Unlike dual variables, quasi-duals admit both positive and negative values at off-optimal operating points, distinguishing constraints that actively oppose cost reduction from those that are merely active or violated. This provides a complementary economic diagnostic that separates geometric proximity to limits from economic relevance.

The framework was extended to probabilistic robustness assessment under both distribution-based and moment-based uncertainty models. To manage combinatorial regime changes under uncertainty, we developed a sequentially pruned tree for mixture models when distributions are known, and an $\alpha$-stressed regime for when only first and second moments are available. These methods enable tractable estimation of violation and release probabilities while remaining consistent with physical regime transitions.

A case study on a model of the Puerto Rican bulk power system demonstrated how the proposed diagnostics can be used to increase situational awareness of system robustness. The sample visualizations selected from the broader set of results illustrated how different screening metrics emphasize complementary aspects of system stress, including contingency-driven violations, directional vulnerability, economic tightness, and probabilistic risk, providing insight beyond feasibility and optimality alone. 

Several directions for future work remain, including extensions to more detailed steady-state voltage control logic, including AVC and $PQ\!\rightarrow\!PV$ transitions, higher-order response mappings, and improved probabilistic quasi-dual estimation. More broadly, the proposed framework provides a systematic means of auditing the robustness of AC-OPF solutions across scenario-based operating conditions, serving as a post-optimal diagnostic layer alongside primal and dual solutions for planners and operators.

\section*{Declaration of Competing Interest}
The authors declare that they have no competing financial interests or personal relationships that could have appeared to influence the work reported in this paper.

\section*{Data Availability}
The~base power flow model and AC OPF solutions used in the case study are available in the following GitHub repository: \url{https://github.com/lauanton/puerto_rico_robustness_case_study}. Accessed on: February 10, 2026.

\section*{Acknowledgments}
The research presented in this manuscript was funded by the MIT Martin Family Society of Fellows for Sustainability. The case study uses a publicly available model of the Puerto Rican system described in \cite{2024_sm_thesis}. The authors thank Rupamathi Jaddivada, Director of Innovation at \emph{SmartGridz}, for~her support and initial guidelines on using the \emph{SmartGridz} AC optimal power flow~solver.

\bibliographystyle{ieeetr}
\bibliography{references}

\end{document}